\documentclass{article}

\usepackage{arxiv}

\usepackage[utf8]{inputenc} 
\usepackage[T1]{fontenc}    
\usepackage{hyperref}       
\usepackage{url}                
\usepackage{booktabs}       
\usepackage{amsfonts}       
\usepackage{nicefrac}       
\usepackage{microtype}      
\usepackage{lipsum}
\usepackage{multirow}
\usepackage{graphicx}
\usepackage{subcaption}
\usepackage[symbol]{footmisc}

\title{Estimating the resilience to natural disasters by using call detail records to analyse the mobility of internally displaced persons}

\author{
  Tracey Li\thanks{tracey.li@flowminder.org} , Jesper Dejby, Maximilian Albert, Linus Bengtsson, V\'eronique Lefebvre\thanks{veronique.lefebvre@flowminder.org} \\
  Flowminder Foundation\\
  17th May, 2019
  }
    
\begin{document}
\maketitle
\vspace{1cm}
\begin{abstract}
We use mobile phone call detail records to estimate the resettlement times of a subset of individuals that have been previously identified to be internally displaced persons (IDPs) following a sudden-onset disaster \cite{li2019}. Four different mobility metrics - two versions of radius of gyration and two versions of entropy - are used to study the behaviour of populations during three disasters - the 2010 earthquake in Haiti, the 2015 Gorkha earthquake in Nepal, and Hurricane Matthew in Haiti in 2016. We characterise the rate at which a disrupted population resettles by the fraction of individuals who remain disrupted each week after the disaster. We find that this rate can be modelled very well as the sum of two exponential decays and observe that the resettling rate for all three disasters is similar, with half the original number of displaced persons having resettled within four to five weeks of the disaster. If the study of further disasters leads to the observation of similar exponential decay rates, then it would imply that the number of IDPs at any time can be inferred from an estimate of the initial number of IDPs immediately following the disaster. Alternatively, the method provides a way to monitor disaster resilience and compare recovery rates across disasters. The method has the advantage that no assumptions need to be made regarding the location or time of resettlement. We also find that metrics that are skewed by long distances, such as the radius of gyration, may not be ideal in this specific context because post-disaster disruption manifests in many individuals as a decrease in long-distance travel and an increase in short-distance travel. A metric that is skewed by long distances can therefore make it appear as though these individuals have recovered earlier than they really have, resulting in an underestimate of the number of people requiring support at a given time. Our results indicate that CDRs can significantly contribute to measuring and predicting displacement durations, distances, and locations of IDPs in post-disaster scenarios. We believe that information and estimates provided by specifically developed CDR analytics, coupled with field data collection and traditional survey methods, can assist the humanitarian response to natural disasters and the subsequent resettlement efforts.

\end{abstract}


\section{Introduction}

Natural and man-made disasters result in millions of people worldwide being displaced each year. The United Nations High Commission for Refugees (UNHCR) reports that over 68 million people in the world are currently displaced \cite{unhcr2017}. Ensuring that adequate support and services are provided for displaced persons is the aim of organisations such as the International Organisation for Migration (IOM). Achieving this requires access to accurate data on how many people have been displaced, where they have been displaced from, and where they have been displaced to, at a given point in time. However, this information is difficult, time-consuming, and expensive to obtain as it is collected via manual surveys that are conducted at known displacement sites. These sites are often large and provide temporary shelters to several thousands of people, making a systematic survey of everyone entering or leaving the site difficult. There may also be many people who are displaced but who do not choose, or are unable to go to organised displacement sites and may be left uncounted \cite{unhcr2012}. There is therefore a need for data collection techniques that can reach beyond the limitations of manual survey methods.

In addition to being required for the immediate response effort, accurate data about displacement are also necessary to further our understanding of how populations behave in times of crisis. Analysing multiple datasets can enable common patterns to be identified. These insights can feed into general models of post-disaster mobility behaviour that can help to predict how a population will respond to a disaster. This can lead to significant improvements in disaster-preparedness efforts.
 
The analysis of mobile phone call detail records (CDR) has been proven to be an effective and valuable way of studying large-scale human mobility after natural disasters \cite{bengtsson2011, lu2012, wilson2016}. These are data that are collected by mobile network operators (MNOs) for billing purposes with one record being generated each time a call or SMS is made or received. The record includes an identifier of the SIM card, the timestamp of the call or SMS, and the location of the cell tower to which the call or SMS was routed (this is usually the tower that is closest to the caller). Assuming that each SIM card is used by a single person, this data enables an individual’s spatio-temporal trajectory to be reconstructed, with the location of the cell tower being taken to be the approximate location of the subscriber. As mobile phone penetration continues to increase in all regions of the globe, including in low- and middle-income countries \cite{gsma2017}, CDR analysis is becoming a method through which the movements of large numbers of people can be studied in a meaningful and efficient way because data for a large number of people are available in near real-time and free of interview bias. 

In previous work \cite{li2019}, we developed a method to analyse CDR data to filter out the subset of a population that we believe to be internally displaced persons (IDPs) who have been displaced as the result of a sudden-onset disaster. We follow on from the work described in \cite{li2019}, using the same datasets and studying the same disasters (Table \ref{tab:disasters}). The subset of data used in this current work is the output of the method developed in \cite{li2019}. As in \cite{li2019}, we would like to highlight that this analysis of CDR data should be interpreted with several caveats in mind, and that CDR-based analyses should be viewed as being complementary to traditional methods rather than as a replacement for them. This is because CDR data are rarely representative of an entire population, especially in low-income countries, since only people who own and use a SIM card are included in the dataset. This typically excludes the youngest and oldest segments of a population, as well as those in the lowest socioeconomic strata. Furthermore, for analysis purposes, it is often only people who call above a minimum frequency that can be included as there are otherwise an insufficient number of data points to perform any meaningful analysis. 

In this study, we focus on one aspect of displaced behaviour, which is the time taken for an IDP to resettle. Resettlement can be considered to be part of an individual’s recovery, which is one aspect of `resilience’ (see e.g. \cite{cutter2008, paton2017}). We develop a method to determine the time at which an individual’s mobility behaviour returns to normal in the aftermath of a disaster and use this as an estimation of the time at which that individual resettled. The method is applied to the three datasets that are summarised in Table \ref{tab:disasters} and described in detail in \cite{li2019}. The results from each dataset are analysed individually and also compared with each other.

\begin{table}
	\caption{Summary details of the studied disasters.}
  	\centering
  	\begin{tabular}{llll}
  	\toprule
    			& Haiti Earthquake & Hurricane Matthew & Nepal Earthquake \\
 	\toprule
	Disaster date &  2010-01-12 & 2016-10-04 & 2015-04-25 \\
	Duration of pre-disaster period & 42 days &182 days & 112 days \\
			& (6 weeks) & (26 weeks) & (16 weeks) \\
	Duration of post-disaster period & 182 days & 153 days & 168 days \\
			& (21 weeks) & (26 weeks) & (24 weeks) \\
	Number of IDPs in analysis & 37,839 & 51,070 & 92,806 \\
 	\bottomrule
  	\end{tabular}
  	\label{tab:disasters}
\end{table}

\section{Method}
\label{sec:method}

We calculate four different mobility metrics - two versions of radius of gyration and two versions of entropy - to describe an individual’s mobility behaviour over the course of the studied time period. These metrics will be described in Section \ref{sec:metrics}. We use each of them to estimate the time at which an individual resettled after being disrupted by a disaster and compare the results obtained by the different metrics.

We characterise an individual’s ‘normal’ behaviour by the average value of a mobility metric in the pre-disaster period. The post-disaster value of the same metric, each week, is then compared with the pre-disaster baseline value. The week on which the value returns to its pre-disaster value or lower is defined to be the ‘resettling date’ of that individual. A key assumption that we make here is that disaster-induced disruption manifests as an increase only (not decrease) in the value of the mobility metric, and that an individual’s behaviour returning to normal is indicative of them resettling. 

The steps of the method are:
\begin{enumerate}
\item Calculate a mobility metric for each week in the dataset.
\item Calculate the mean weekly value of the metric, aggregating over all weeks in the pre-disaster period. This is the benchmark value to define someone's `normal' behaviour.
\item For the post-disaster period, starting on the day of the disaster, calculate the four-week rolling mean value of the metric for each week, using a window to the right of the date.
\item Find the first week in the post-disaster period during which the four-week mean value is equal to or lower than the benchmark pre-disaster period. Use this as the resettling date.
\end{enumerate}

The use of four weeks to calculate the rolling mean is a somewhat arbitrarily chosen value - it is the minimum value that we consider to be adequate for our purposes - but have verified that the results and conclusions that are presented in this work are not strongly affected by varying this parameter within a reasonable range.

This work is novel and experimental, both because of the method used to identify IDPs \cite{li2019}, and because the mobility metrics used are either unconventionally defined or have been adapted specifically for this dataset and use case. It is therefore important to verify that any results we observe are actually manifestations of real behaviour and are not just artefacts of the method. We perform this validation by defining a control group that consists of a set of randomly chosen subscribers who have not been identified to be IDPs but that were part of the `frequent caller' subset defined in Section 3 of \cite{li2019}. It should be noted, however, that it is likely that several individuals in the control group actually are IDPs although they have not been labelled as such, since we were deliberately conservative in our method for labelling IDPs - this was to ensure that we have a very pure sample with a low false positive rate. The control group contains the same number of individuals as the IDP group and we perform all analyses on both the set of IDPs that were selected out by the method in \cite{li2019}, and on the control group. By comparing the results we can distinguish between any fake effects that are observed in both groups from real effects that are observed only in the IDP group.

\subsection{Mobility metrics}
\label{sec:metrics}
Several metrics have been developed in the literature to study human mobility using mobile phone data. These include average distance travelled \cite{lu2012, rubio2010}, radius of gyration \cite{lu2012, gonzalez2008, song2010}, area of influence \cite{rubio2010}, moment of inertia \cite{gonzalez2008}, and entropy \cite{lu2012, song2010}. These measures are sensitive to different aspects of mobility, such as distance travelled, frequency of travel, number of distinct locations visited, frequency of visits to each location, and regularity of visits. The utility of each metric is therefore dependent on the context.

We have chosen to use two of the most commonly-used metrics - radius of gyration (as used, for example, in \cite{lu2012, gonzalez2008, song2010}) and entropy, as in \cite{lu2012, song2010}. The former is a measure of the linear space occupied by an individual, whereas the latter is a measure of the orderliness, or predictability, of an individual’s trajectory. We have used both a standard definition and a modified definition for each of the two metrics, all of which are described in the subsequent sections. We compare the results obtained by these metrics.  

The metrics have been adapted to be calculated from the `distance curves’ that were described in detail in Section 3 of \cite{li2019}. The distance curve for an individual phone subscriber is a time series formed of the distances between the daily location assigned to them each day and their pre-disaster home location (see \cite{li2019} for method details). Values for dates on which there are no data because no calls were recorded are imputed, using the method described in Section 3 of \cite{li2019}. The mobility metrics were calculated by aggregating over each week of data, to obtain weekly values. We note that this is a short time period that only contains seven data points which makes the data very sensitive to noise. We choose to do this because we want to be sensitive to sudden shocks, to which short time periods with a small number of data points are very sensitive.

\subsubsection{Radius of gyration (RoG) - standard definition}
The radius of gyration (RoG), as used in e.g. \cite{lu2012}, is typically defined as the standard deviation of the distances between each visited location and the centroid, over a specified time period: $\mbox{RoG} = \sqrt{\frac{1}{n}\Sigma_{i=1}^{n}(r_{i}-r_{c}) }$ where $n$ is the length of the time period. The centroid, $r_{c}$, is calculated as the mean location, weighted by the frequency of visits to each location. The RoG depends on the distances between locations and the number of visits to each location. As standard deviations are easily skewed by anomalously high values, long-distance trips are highly weighted in this metric.

For reasons of simplicity and efficiency, we have used an approximation of this metric which can be calculated from the distance curves we produced in \cite{li2019}. Additionally, because we are using only a small number of points (seven) for each calculation, we calculate the mean absolute deviation instead of the standard deviation. The steps of the calculation are as follows:
\begin{enumerate}
\item For each week, calculate the centroid as the mean distance from home location during that week.
\item Calculate the modulus of the difference between each daily location and the corresponding week’s centroid.
\item Calculate the mean of the absolute differences for each week.
\end{enumerate}

We note that the method described above is identical to the true radius of gyration only in the scenario where all visited locations lie along a one-dimensional line. This is clearly never the case. Also, we will point out that the `distance curves' calculated in \cite{li2019} consist only of a single location for each day (the one nearest to the home location) and so many locations are not included. However, as we are primarily interested in determining \textit{changes} in the metric between the pre- and post-disaster periods, we believe that our approximation is sufficient and has the benefit of a reduction in complexity.

\subsubsection{Modified radius of gyration - logarithmic radius of gyration (LRoG)}
From the initial analysis of our results in \cite{li2019}, we observe that many displacements occur over very short distances of less than a few kilometres. Therefore, it is important to ensure that we are sensitive to changes in the frequency of these short-distance trips as well as long-distance trips. We have therefore created a modified definition of the RoG that is more sensitive to short trips, which is calculated as follows:
\begin{enumerate}
\item For each week, calculate the centroid as the mean distance from home location during that week.
\item Calculate the logarithm of the modulus of the difference between each daily location and the corresponding week’s centroid. Using the logarithm ensures that short-distances are weighted more highly than if the linear distance is used.
\item Calculate the mean of the logarithmic differences for each week.
\end{enumerate}

We will refer to this modified metric as the \textit{logarithmic radius of gyration}, abbreviated to LRoG.

\subsubsection{Entropy - original definitions}
The entropy definitions used in e.g. \cite{lu2012} were developed in \cite{song2010}. These definitions are particularly suited to quantifying the predictability of someone’s movements. Several versions of entropy are defined \cite{song2010}, with formulae provided in Table \ref{tab:entropy}:
\begin{itemize}
\item The true entropy accounts for the relative frequency of visited locations and the order in which they are visited. It is calculated by enumerating the probabilities of all subsequences appearing in a time series. 
\item The temporal-uncorrelated entropy depends only on the frequency at which locations are visited, but not the order in which they are visited. 
\item The random entropy is the simplest definition, and depends only on the number of distinct locations that are visited, independent of frequency or ordering.
\end{itemize}

We have chosen to include the temporal-uncorrelated entropy, S\textsuperscript{uncorr}, in our study. This metric can be calculated relatively simply from the distance curves described in \cite{li2019} by counting the number of days spent at each daily location.

\subsubsection{Modified entropy definition - step entropy (S\textsuperscript{step})}
The definitions of `stability’ and `displacement’ that we used in \cite{li2019} are based on measuring changes in `constant’ daily locations, or `stay locations'. They are fundamental to our methodology. For consistency and continuity, we would like to use a mobility metric that is based on these definitions and, specifically, is sensitive to changes in location. Staying for a short duration at a location and therefore frequently changing location should result in a higher score than staying for long durations at each location. The number of distinct locations visited, and the number of distinct visits to each location, therefore need to be taken into account. Additionally, the metric should be independent of distance so that short travel distances are counted as equally as long distances.

We have therefore created another entropy-based metric that satisfies these properties. The metric is based on the definition of Boltzmann’s entropy in statistical thermodynamics. This is a particular instance of Gibbs entropy which is identical to the definition of Shannon entropy used in information theory, on which the entropy definitions in \cite{song2010} are based. We will refer to this new metric as step entropy, S\textsuperscript{step}, following on from the step detection method we implemented in \cite{li2019}. The metric is calculated as follows:
\begin{enumerate}
\item Reduce consecutive visits to the same location to a single visit e.g. AAABA $\rightarrow$ ABA.
\item Calculate the number of distinct permutations of the sequence obtained above: $P =  n!/(m_{1}! m_{2}! m_{3}! ... )$ where $n$ is the total number of visits in the sequence and $m_i$ is the number of times the location $i$ appears in the sequence.
\item Take the logarithm of P and define this to be the entropy: S\textsuperscript{step} = $\mbox{log}[n!/(m_{1}! m_{2}! m_{3}! ... )] = log(n!) - log(m_{1}!) - log(m_{2}!) - log(m_{3}!) - …$ .
\end{enumerate}
The metric is a measure of disorder, in terms of the number of ways that a sequence of locations can be ordered. This is influenced both by the number of distinct locations that are visited, and the number of visits to each location, after consecutive visits to the same location have been contracted as a single visit or `stay'. A large value of the metric indicates low predictability of movements and high mobility. The metric can be calculated very simply from the distance curve after it has been processed into a piecewise-constant signal (see `Method' section of \cite{li2019}).

\subsubsection{Comparison of entropy definitions}
Table \ref{tab:entropy} shows a summary of the definitions of entropy discussed above. 
\begin{table}[h!]
	\caption{Entropy definitions.}
  	\centering
  	\begin{tabular}{ll}
  	\toprule
    	Definition	&	Behaviours to which it is sensitive	 \\
 	\toprule
	S\textsuperscript{random} = log($L_{i}$)	& Number of distinct locations.	\\
	where $L_{i}$ = number of distinct locations.	&					\\
	\midrule
	S\textsuperscript{uncorr} = $-\Sigma_{k}p_{k}\mbox{log}(p_{k})$	& Number of distinct locations.	\\
	where $p_{k}$ is the frequency of visiting location $k$.	& Frequency of visits to each location.	\\
	\midrule
	S\textsuperscript{true} = $\Sigma_{X_{i}} p(X_{i})\mbox{log}(p(X_{i}))$	& Number of distinct locations.	\\
	where $p(X_{i})$ is the probability of finding a 				& Frequency of visits to each location.	\\
	subsequence $X_{i}$ in the time series.					& Ordering of visits to each location. \\
	\midrule
	S\textsuperscript{step} = $\mbox{log}(n! / \Pi_{i}m_{i}!)$	& Number of distinct locations.	\\
	where $n$ is the total number of stays and $m_{i}$  & Frequency of changes of location. \\
	is the number of stays at location $i$.		&	\\
 	\bottomrule
  	\end{tabular}
  	\label{tab:entropy}
\end{table}

\section{Results}

We present our results in the form of `decay curves’ that show how the fraction of IDPs who have not yet resettled decreases over time. These curves depict the rate at which people return to normal, after they have experienced a shock that disrupts their behaviour and increases their level of mobility. The time taken for them to return to normal is what we refer to as the `resettlement time' from now on.

\subsection{Resettlement `decay curves'}
\label{sec:decays}

\begin{centering}
\begin{figure}[h!]
  \begin{subfigure}{1.0\linewidth}
    \centering\includegraphics[width=0.65\linewidth]{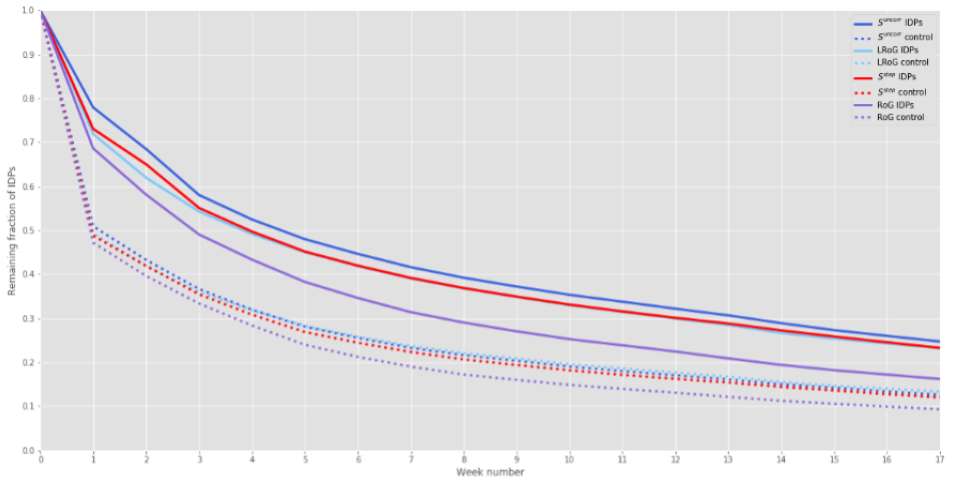}
    \caption{Haiti earthquake.}
  \end{subfigure}
  \begin{subfigure}{1.0\linewidth}
    \centering\includegraphics[width=0.65\linewidth]{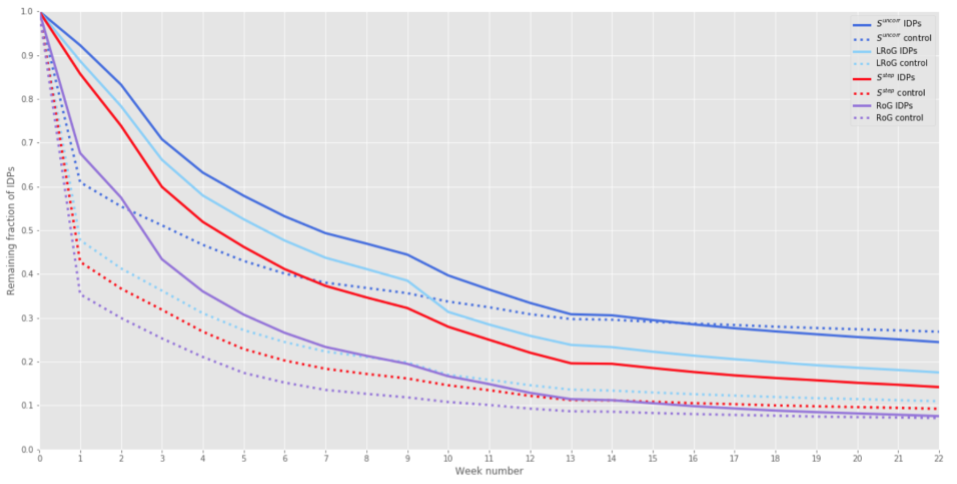}
    \caption{Hurricane Matthew.}
  \end{subfigure}
  \begin{subfigure}{1.0\linewidth}
    \centering\includegraphics[width=0.65\linewidth]{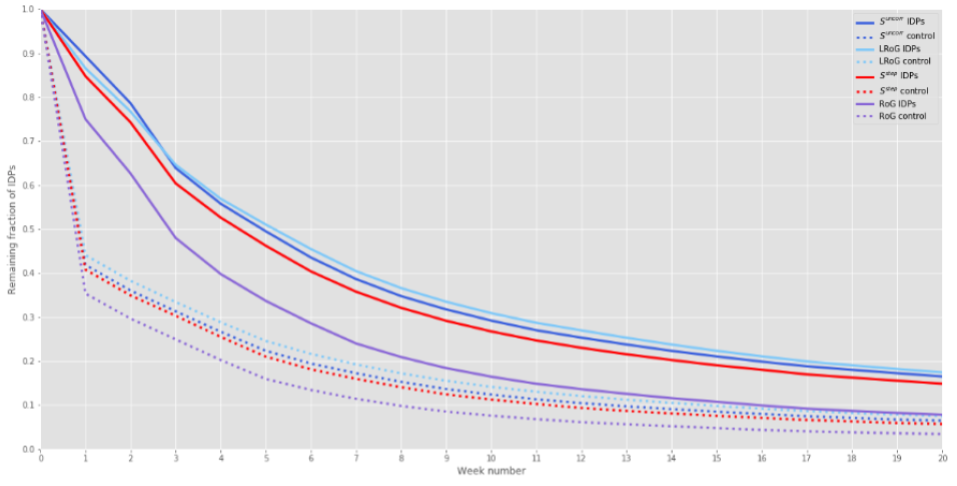}
    \caption{Nepal earthquake.}
  \end{subfigure}
\caption{Resettlement decay curves as calculated from all four mobility metrics. Results for the IDP group are shown by solid lines, and results for the control group are shown by dotted lines.}
\label{fig:decay}
\end{figure}
\end{centering}

The decay curves for the three disasters are shown in Figures \ref{fig:decay}. We show the results obtained from all four mobility metrics, for both the control and IDP groups, for each disaster. The horizontal axis shows the number of weeks (seven day periods) after the disaster, where weeks are defined to start on the same weekday as the disaster. Week 0 denotes the seven-day period beginning on the date of the disaster, during which all IDPs are displaced. The fraction of IDPs remaining at week $n$ is the fraction of IDPs that have not yet recovered $7n$ days after the disaster.

As a verification that the method works as expected, we have performed a similar analysis on the pre-disaster period as the post-disaster period; that is that we perform the same steps 1 and 2 of the method described in Section \ref{sec:method}, but for step 3 we start on the first day of the dataset and calculate the four-week rolling mean for each week in the pre-disaster period. In step 4, we find the first week in the pre-disaster period in which the four-week mean is equal to or lower than the benchmark value calculated in step 2. (This is analogous to training a model on a portion of the data, and then evaluating how the model performs on that same training data). These pre-disaster decay curves are shown in Appendix \ref{app:decay}. We find that the recovery rates are similar for the control and IDP groups, which is the expected behaviour during an undisrupted period. The decay rate is slightly faster for the control group than the IDP group, which indicates that the mobility of the IDP group is inherently more volatile than that of the control group even during a ‘normal’ period. In the post-disaster period however, there is a very obvious difference between the IDP and control groups, with the IDP group recovering much more slowly than the control group, as expected (Figures \ref{fig:decay}). This gives us confidence that this method can be used to assess the behaviour of IDPs and study how their behaviour deviates from normal in the aftermath of a disaster.

We observe that for all three disasters, the curve for RoG decays much faster than for the other metrics. This indicates that the mobility characteristics to which RoG is sensitive (e.g. long-distance travel patterns) return to normal faster than other mobility characteristics to which the other metrics are sensitive (e.g. very short-distance travel patterns, and number of distinct locations visited).

In all subsequent figures, for simplicity and clarity, we have chosen to show the curves for the step entropy metric only, with the choice of this metric being discussed later in Section \ref{sec:compare_metrics}.

\subsection{Exponential mixture model}
\label{sec:exp_fit}
\begin{centering}
\begin{figure}
  \begin{subfigure}{1.0\linewidth}
    \centering\includegraphics[width=0.65\linewidth]{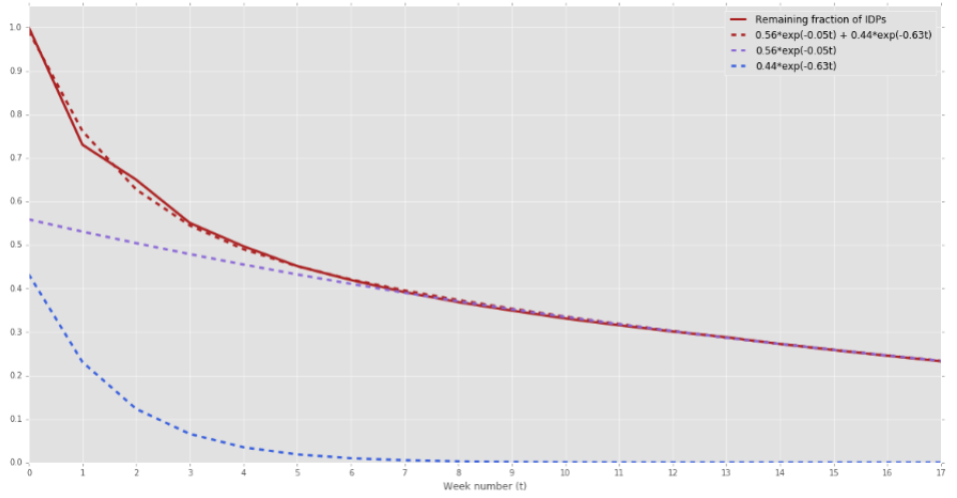}
    \caption{Haiti earthquake.}
  \end{subfigure}
  \begin{subfigure}{1.0\linewidth}
    \centering\includegraphics[width=0.65\linewidth]{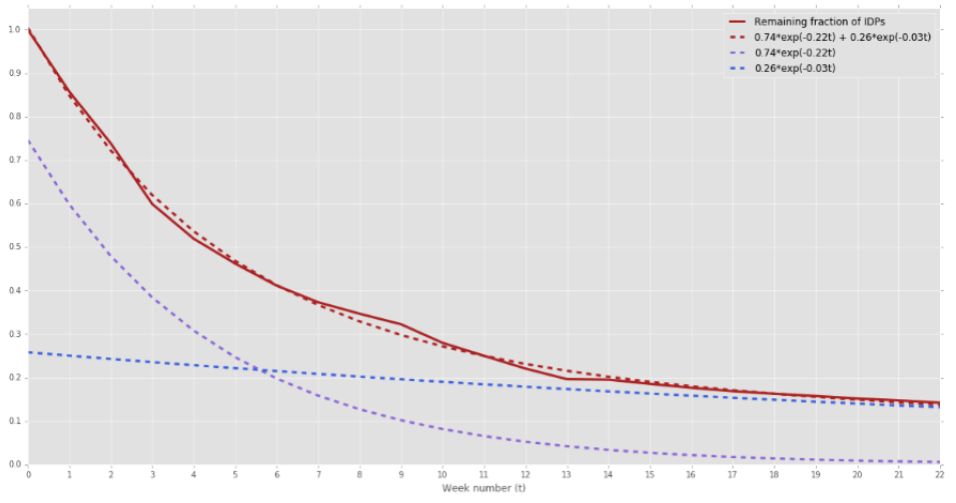}
    \caption{Hurricane Matthew.}
  \end{subfigure}
  \begin{subfigure}{1.0\linewidth}
    \centering\includegraphics[width=0.65\linewidth]{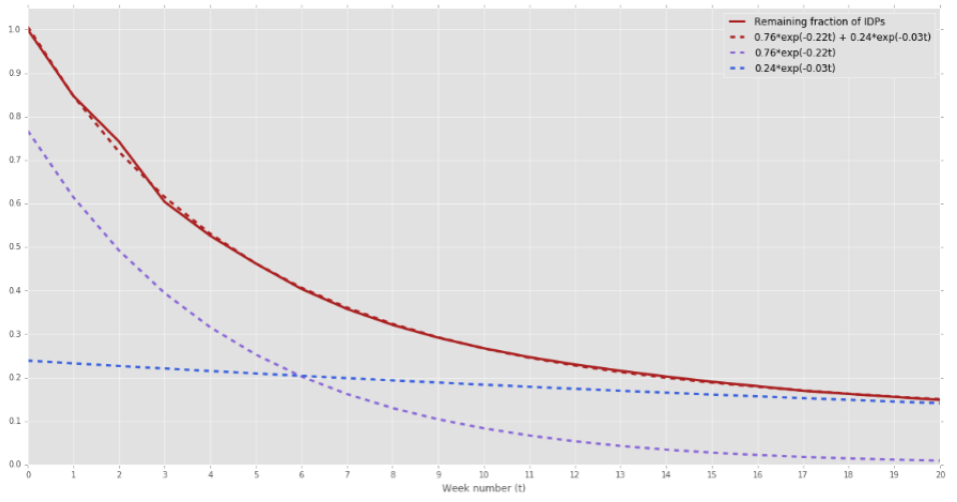}
    \caption{Nepal earthquake.}
  \end{subfigure}
\caption{Resettlement decay curves (solid red line) together with a fit to the sum of two exponentials (dashed red line), and the individual exponential components - one fast decay (dashed purple line) and one slow decay (dashed blue line).}
\label{fig:exp_fit}
\end{figure}
\end{centering}

\begin{figure}
  \centering\includegraphics[width=0.65\linewidth]{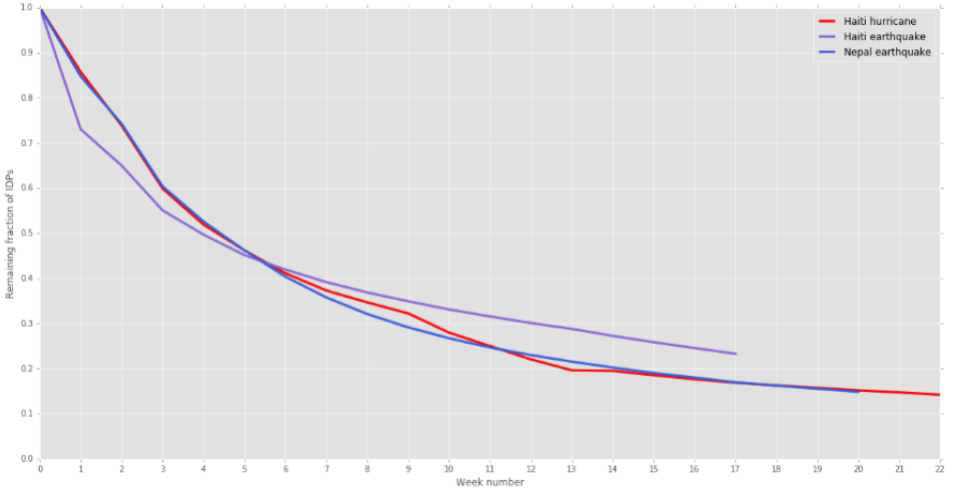}
  \caption{Resettlement decay curves for all three disasters.}
  \label{fig:decay_all}
\end{figure}

In Figure \ref{fig:exp_fit} we show the step entropy decay curves for each disaster (solid red line) together with a fit to the sum of two exponential curves, parameterised as $f(t) = \alpha_{1}\mbox{exp}(-\beta_{1}t) + \alpha_{2}\mbox{exp}(-\beta_{2}t)$ (dashed red line), and the two individual exponential curves that compose the sum. The values of the parameters $\alpha_{1}$, $\alpha_{2}$, $\beta_{1}$, and $\beta_{2}$ for each disaster are shown in Table \ref{tab:parameters}, and the number in parentheses after each value is the 1-standard deviation error on the parameter. All three curves fit very well to the sum of two exponential decays - one ‘fast’ decay and one ‘slow’ decay. We interpret this combination of decays as corresponding to two distinct groups of IDPs - one group that recovers relatively quickly, and a second group that recovers more slowly. We observe that the decay curves for two of the disasters - Hurricane Matthew and the Nepal earthquake - are extremely similar. For the third disaster, the recovery rate is faster immediately after the disaster then decreases more rapidly. However, it is still similar to the other two curves. This can be seen most clearly in Figure \ref{fig:decay_all} where we show the decay curves for all three disasters plotted together. We summarise these curves in Table \ref{tab:halflife} by showing the amount of time it takes for the number of remaining IDPs to decrease to a half, and then a quarter, of the original number. The shorter this amount of time, the faster the rate of recovery. 

In further work, it would be valuable to explore whether the recovery rates of other disasters are also similar to the ones we have observed here, as well as to investigate what factors influence the values of the model parameters. This would further our understanding of what affects disaster recovery times, providing the possibility to predict these times in future disasters and to put in place interventions that reduce these times.

\begin{table}[h!]
	\caption{Best-fit parameter values when fitting the fraction of IDPs that remain displaced at time $t$ to the model $f(t) = \alpha_{1}\mbox{exp}(-\beta_{1}t) + \alpha_{2}\mbox{exp}(-\beta_{2}t)$.}
  	\centering
  	\begin{tabular}{llll}
  	\toprule
    		&	Haiti Earthquake & Hurricane Matthew & Nepal Earthquake \\
 	\toprule
	$\alpha_{1}$ & 0.43 (0.02) & 0.75 (0.07) & 0.76 (0.001) \\
	$\alpha_{2}$ & 0.56 (0.02) & 0.26 (0.07) & 0.24 (0.001) \\
	$\beta_{1}$   & 0.63 (0.05) & 0.22 (0.02) & 0.22 (0.0001) \\
	$\beta_{2}$   & 0.05 (0.002) & 0.03 (0.01) & 0.03 (0.00001) \\
	\bottomrule
  	\end{tabular}
  	\label{tab:parameters}
\end{table}

\begin{table}
	\caption{Number of weeks taken for the number of IDPs that remain displaced to decrease to $\frac{1}{2}$ and $\frac{1}{4}$ of the original number.}
  	\centering
  	\begin{tabular}{llll}
  	\toprule
    		&	Haiti Earthquake & Hurricane Matthew & Nepal Earthquake \\
 	\toprule
	$\frac{1}{2}$ & 4 & 5 & 5 \\
	$\frac{1}{4}$ & 17 & 12 &12 \\
	\bottomrule
  	\end{tabular}
  	\label{tab:halflife}
\end{table}

\subsection{Comparing recovery times between regions}

\begin{centering}
\begin{figure}[h!]
  \begin{subfigure}{1.0\linewidth}
    \centering\includegraphics[width=0.65\linewidth]{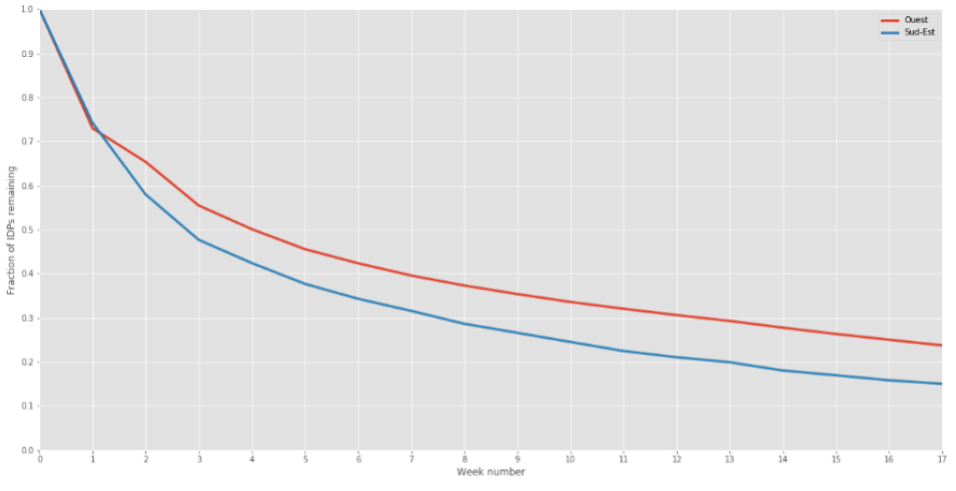}
    \caption{Haiti earthquake.}
  \end{subfigure}
  \begin{subfigure}{1.0\linewidth}
    \centering\includegraphics[width=0.65\linewidth]{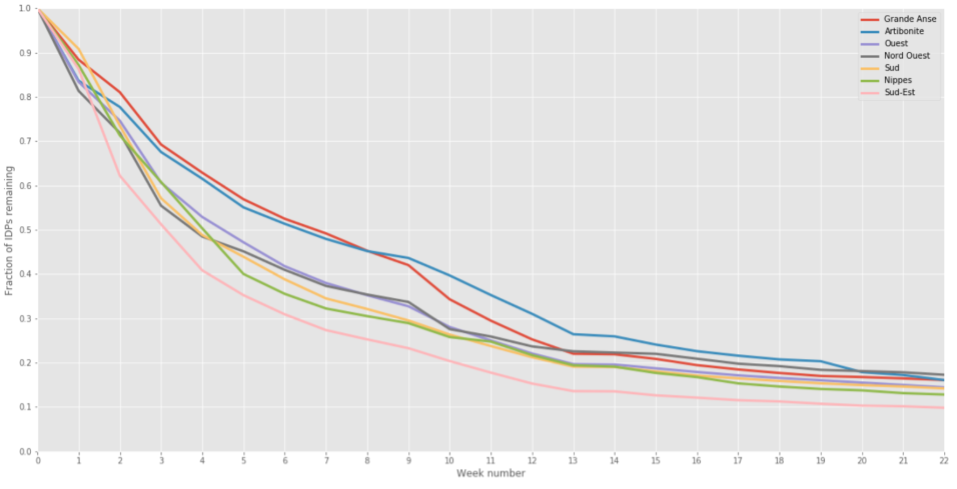}
    \caption{Hurricane Matthew.}
  \end{subfigure}
  \begin{subfigure}{1.0\linewidth}
    \centering\includegraphics[width=0.65\linewidth]{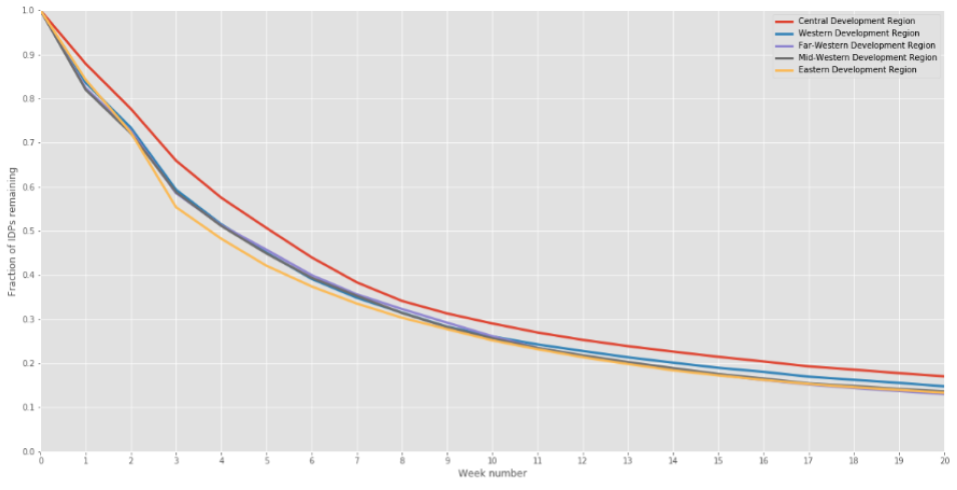}
    \caption{Nepal earthquake.}
  \end{subfigure}
\caption{Resettlement decay curves for each administrative level 1 region}
\label{fig:admin1}
\end{figure}
\end{centering}

In Figure \ref{fig:admin1} we show similar decay curves to those in Figure \ref{fig:decay}, but with one curve for each administrative level 1 region. Qualitatively, these results are consistent with what would be expected in terms of the impact of the disaster on each region - the regions that were reported to be the most severely affected exhibit the slowest recovery rates (Grande Anse for Hurricane Matthew, and Central for the Nepal earthquake).

\subsection{Categorising IDPs into returnees, and those who resettle elsewhere}
For the individuals for whom a resettling date can be determined within the time period of the dataset, their resettlement location is determined as the modal daily location (the location on which they are seen on the most unique days) during the period starting from their resettlement date until the end of the dataset. This is the same method that is used to determine the home location in the pre-disaster period, which is described in detail in \cite{li2019}. This group can then be divided into the individuals that resettled at their pre-disaster home location, and those that resettled elsewhere.

We use the following method to categorise an individual as a returnee (resettled at pre-disaster home location), resettled elsewhere (resettled at a different location to their pre-disaster home location), or unstable (not resettled by the end of the dataset):
\begin{enumerate}
\item Assume that the resettlement date, calculated using the mobility metrics described Section \ref{sec:metrics}, is the date at which an IDP returned to a stable location. If there is no resettlement date for an IDP, classify them as `unstable’.
\item For the IDPs that do have a resettlement date, determine their resettlement location for the duration of time from their resettlement date until the end of the dataset, using the same method that is used to determine the pre-disaster reference location. 
\item If the resettlement location is the same as the pre-disaster reference location, classify the IDP as a `returnee’. Otherwise classify them as `resettled elsewhere'. 
\end{enumerate}

\begin{centering}
\begin{figure}
  \begin{subfigure}{1.0\linewidth}
    \centering\includegraphics[width=0.65\linewidth]{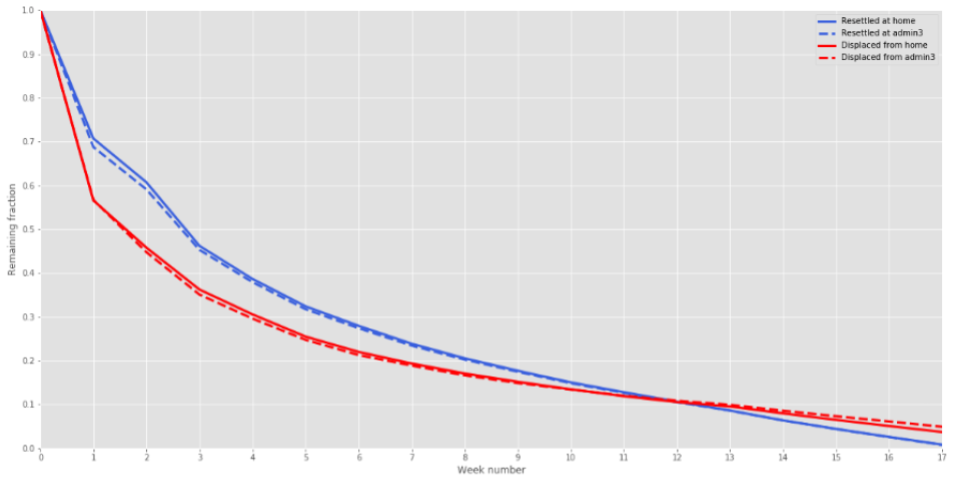}
    \caption{Haiti earthquake.}
  \end{subfigure}
  \begin{subfigure}{1.0\linewidth}
    \centering\includegraphics[width=0.65\linewidth]{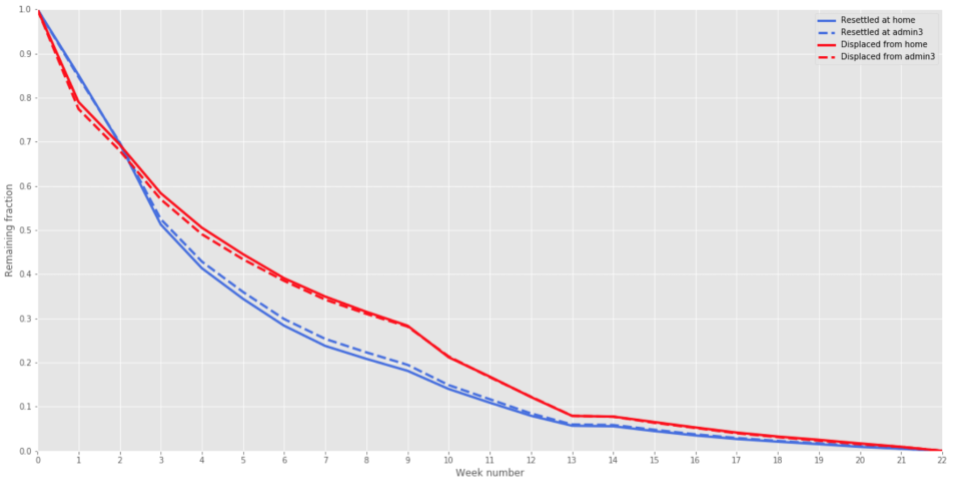}
    \caption{Hurricane Matthew.}
  \end{subfigure}
  \begin{subfigure}{1.0\linewidth}
    \centering\includegraphics[width=0.65\linewidth]{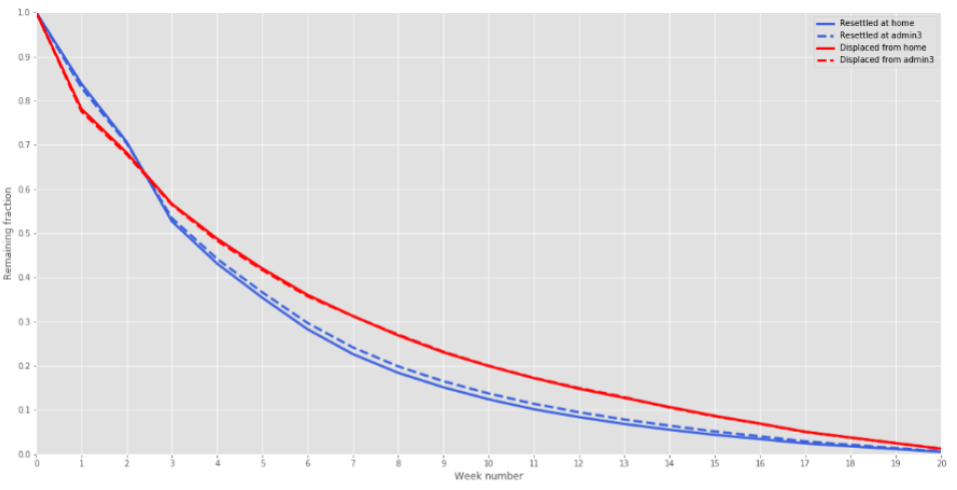}
    \caption{Nepal earthquake.}
  \end{subfigure}
\caption{Resettlement decay curves separated into IDPs that resettle at home - either exact home location or same admin 3 region (blue lines), or at another location (red lines).}
\label{fig:returns}
\end{figure}
\end{centering}

Figure \ref{fig:returns} shows similar decay curves to Figure \ref{fig:decay}, but normalised according to the number of people who have resettled at home, versus a different location. We consider two versions of `home’ - one being a return to exactly the same location as the pre-disaster home location, and the second being a return to the same administrative level 3 region. The fractions of IDPs that constitute each of these groups is shown in Table \ref{tab:resettlement_fractions}. We compare the fraction of IDPs that have resettled after 17 weeks because this is the duration of the dataset with the shortest post-disaster duration (Haiti earthquake), minus four weeks (the length of the rolling mean window used in the method described in Section \ref{sec:method}).

Although there is some difference, the resettlement rates are quite similar for both groups, for all three disasters. In other words, the rate at which people resettle at their original home location is similar to the rate at which people resettle at a new location. One possible interpretation of this is that the time required to repair damage to an existing home is roughly the same as the time that is needed for an individual to find a new home or, more generally, the time it takes to resume `normal' activities such as work and other economic activities. 

\begin{table}
	\caption{Proportions of IDPs that resettle at their pre-disaster home location, and at the same administrative level 3 region as their pre-disaster home location.}
  	\centering
  	\begin{tabular}{llll}
  	\toprule
    		&	Haiti Earthquake & Hurricane Matthew & Nepal Earthquake \\
 	\toprule
	Fraction of IDPs that have		& 0.77 & 0.88 & 0.84 \\
	resettled anywhere by week 17 & & & \\
	\midrule
	Fraction of IDPs that have & 0.57 & 0.73 & 0.78 \\
	resettled at same admin 3 & & & \\
	region by week 17 & & & \\
	\midrule
	Fraction of IDPs that have & 0.49 & 0.63 & 0.60 \\
	resettled at exact home & & & \\
	location by week 17 & & & \\
	\bottomrule
  	\end{tabular}
  	\label{tab:resettlement_fractions}
\end{table}

\section{Discussion}

\subsection{Misidentification of responders as `IDPs'}
It is likely that the subset of subscribers we have identified as IDPs, using the method described in \cite{li2019}, may include local responders to the disaster. For Hurricane Matthew and the Nepal earthquake we observed that roughly 5\% of all displacements originated from areas that were least affected by the disasters, with the destination being to the most affected areas. This proportion increased slightly over time (from 4\% in week 1 to 6\% after week 6) for Nepal, and increased after two weeks before decreasing in the case of Hurricane Matthew. This suggests that people who move from the least affected areas to the most affected areas (possible responders) may remain away or `unstable' for slightly longer than true IDPs. There may also be responders that originate from severely affected areas, who we will not be able to distinguish. In future work, we plan to look at decays based on the disaster intensities of the home and displaced locations. In particular, we will focus on explaining the observed decay rates that we have found can be modelled very well as the sum of exponential functions (Section \ref{sec:exp_fit}). Local responders may influence these decay rates, as well as people whose home was not damaged but who left as a precaution, and then returned almost immediately. 

\subsection{Mobility metrics}
In this section, we briefly compare and evaluate the performance of each of the mobility metrics used in this study. Further analysis and discussion is provided in Appendix \ref{app:pre_post_mobility}.

\subsubsection{Limitations to the use of radius of gyration}
We have found that, by itself, our implementation of RoG, as described in Section \ref{sec:metrics}, is not the most suitable metric for the purpose of assessing when behaviour returns to normality in the aftermath of a disaster. This is because a disaster can alter people’s mobility behaviour in different ways. For some individuals, disruption is exhibited as an increase in the amount of long-distance travel, which is captured well by RoG. However, for many other individuals, disruption occurs as a decrease in long-distance travel, and an increase in the amount of short-distance travel. This can cause the RoG for these individuals to be lower than normal and can therefore lead to the misinterpretation that they have not been affected.

To illustrate this, we examine the pre- and post-disaster behaviours of a test and a control group, where the test group consists of the IDPs for whom the RoG metric has indicated a resettlement date that is earlier than the date obtained by any of the other metrics. The control group consists of the remaining IDPs. 

\begin{centering}
\begin{figure}
  \begin{subfigure}{1.0\linewidth}
    \includegraphics[width=0.5\linewidth]{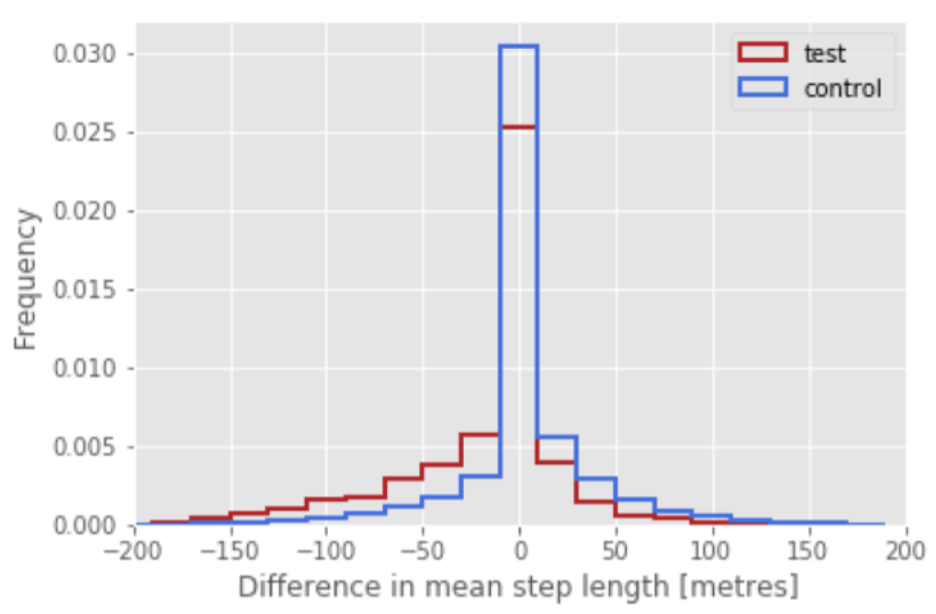}
    \includegraphics[width=0.5\linewidth]{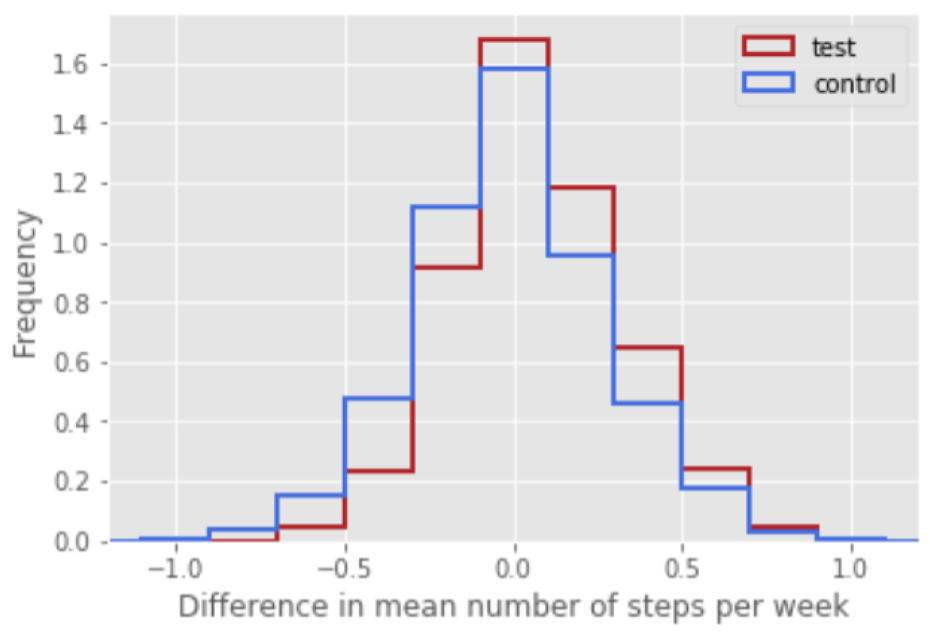}
    \caption{Haiti earthquake.}
  \end{subfigure} \\
  \begin{subfigure}{1.0\linewidth}
    \includegraphics[width=0.5\linewidth]{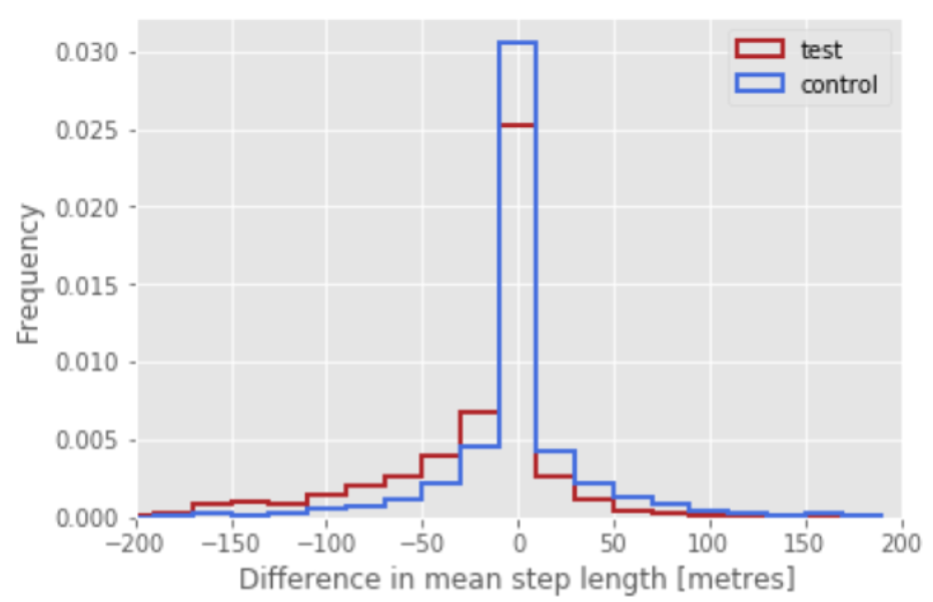}
    \includegraphics[width=0.5\linewidth]{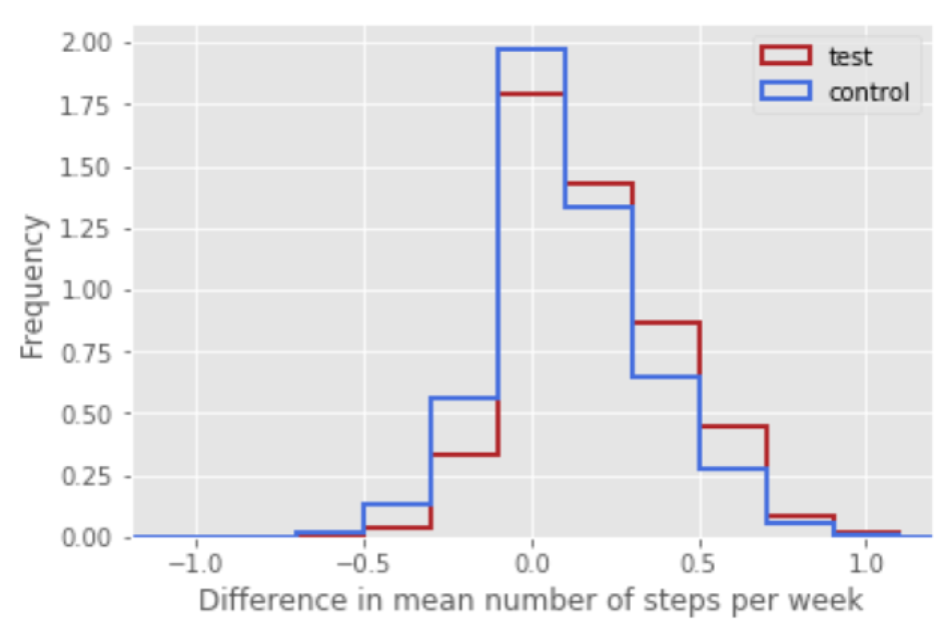}
    \caption{Hurricane Matthew.}
  \end{subfigure} \\
  \begin{subfigure}{1.0\linewidth}
    \includegraphics[width=0.5\linewidth]{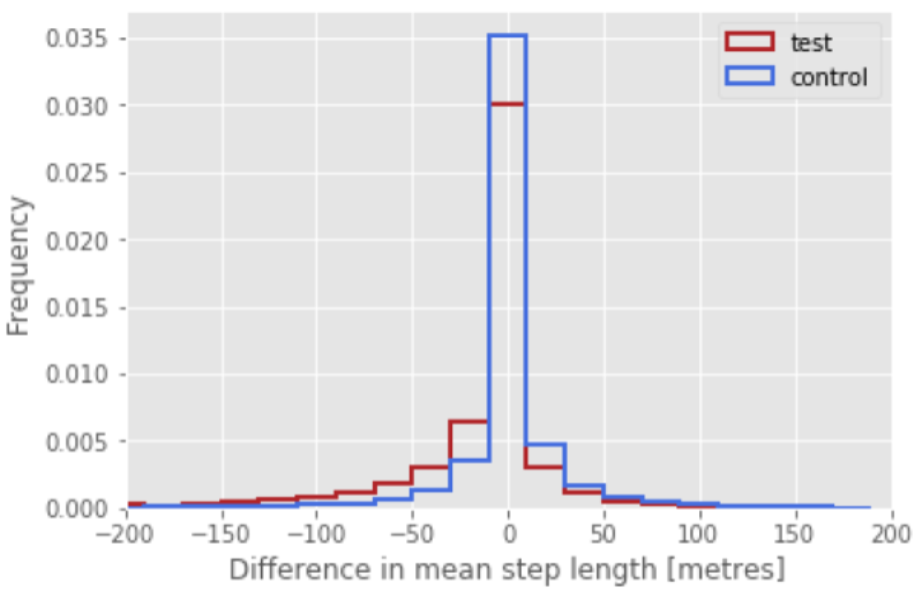}
    \includegraphics[width=0.5\linewidth]{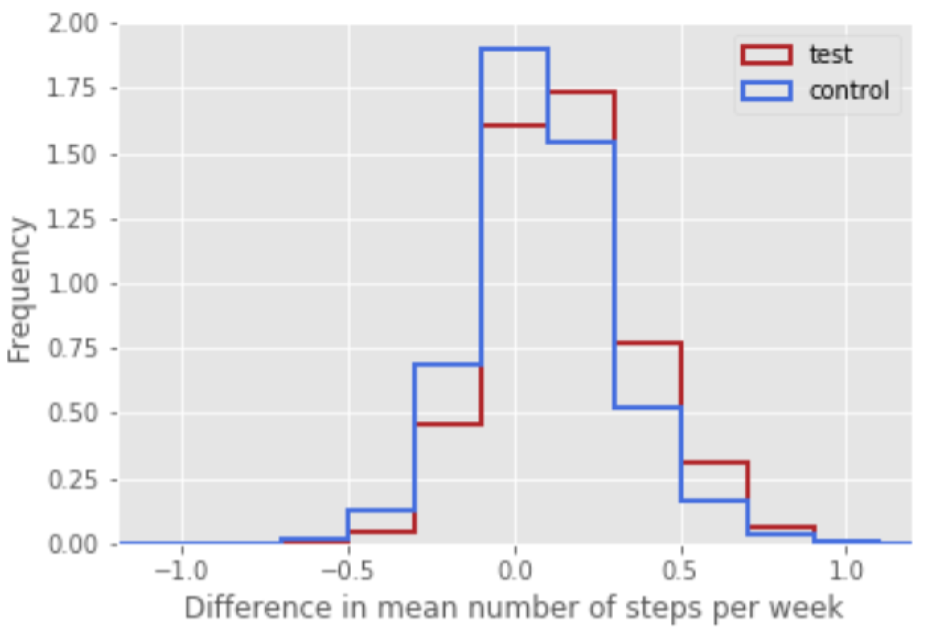}
    \caption{Nepal earthquake.}
  \end{subfigure}
\caption{Distributions of differences between pre- and post-disaster periods in mean step length (left column), and mean number of steps per week (right column) for test and control groups.}
\label{fig:step_differences}
\end{figure}
\end{centering}

To compare the behaviour of these groups, we analyse the piecewise-constant signal with which we are modelling each individual’s movement trajectory (see Section 3 of \cite{li2019}). This is a smoothed and simplified version of the full trajectory that is composed of a sequence of `stay locations', represented by constant ‘pieces’, each one of which represents a stay at a single location for a number of consecutive days. These pieces are joined by ‘steps’, which represent the distance between consecutive locations. We calculate the difference between the mean step distances (and have verified that similar results are obtained by using the median distance) in the post-disaster and pre-disaster periods for both the test and control groups. The distributions are shown in Figure \ref{fig:step_differences}, from which we see that the distribution of differences for the test group are visibly skewed towards the negative side. We have calculated the z-scores for both groups, shown in Table \ref{tab:z_step_length}, and find that this difference is significant for the test group, but that there is no significant change for the control group. We also calculate the z-score to compare the distributions of differences of the test and control groups, and find that these distributions are significantly different, with the test group having a lower mean than the control group. 

We have also calculated the ratio of the mean number of steps made per week in the post-disaster to the pre-disaster periods, also shown in Figure \ref{fig:step_differences}. The skew towards values > 1 shows that more steps were made each week after the disaster than before, indicating that the frequency of travel increased for both groups. However, it can be seen that the test group’s frequency of travel increased more than the control group, and the z-value for this difference is significant (Table \ref{tab:z_num_steps}). We can then conclude that the individuals in the test group typically decrease their travel distance, but increase their travel frequency, more than the individuals in the control group. Therefore, using only RoG can lead to the conclusion that the behaviour of many individuals has returned to normal before it actually has, because of the metric’s sensitivity bias towards long-distance travel. 

\begin{table}
	\caption{z-scores to test differences between mean step length of pre-disaster and post-disaster behaviour.}
  	\centering
  	\begin{tabular}{llll}
  	\toprule
    		&	Haiti Earthquake & Hurricane Matthew & Nepal Earthquake \\
 	\toprule
	Control group		& 20.4 & -6.08 & 4.60 \\
	post- vs pre-disaster & & & \\
	\midrule
	Test group			& -33.1 & -39.2 & -34.7 \\
	post- vs pre-disaster & & & \\
	\midrule
	Test vs control group 	& -43.4 & -41.4 & -45.5 \\
	(post-disaster - pre-disaster)	& & & \\
	\bottomrule
  	\end{tabular}
  	\label{tab:z_step_length}
\end{table}

\begin{table}
	\caption{z-scores to test differences between mean number of steps of pre-disaster and post-disaster behaviour.}
  	\centering
  	\begin{tabular}{llll}
  	\toprule
    		&	Haiti Earthquake & Hurricane Matthew & Nepal Earthquake \\
 	\toprule
	Control group		& -46.0 & 22.2 & 40.2 \\
	post- vs pre-disaster & & & \\
	\midrule
	Test group			& 0.25 & 50.3 & 67.8 \\
	post- vs pre-disaster & & & \\
	\midrule
	Test vs control group 	& 29.0 & 25.2 & 40.0 \\
	(post-disaster - pre-disaster)	& & & \\
	\bottomrule
  	\end{tabular}
  	\label{tab:z_num_steps}
\end{table}

\subsubsection{Comparison of mobility metrics}
\label{sec:compare_metrics}

We chose to display all results after Section \ref{sec:decays} using a single metric only, for simplicity. We chose step entropy for the following reasons: the primary reason is that this metric is consistent with the principles of our IDP detection method in \cite{li2019}, which is based on looking for changes in `stay locations'. The second reason is because the step entropy curve lies is the middle of the curves for the other metrics, for all three disaster datasets, and so it can be considered to be an approximation of the average of all the metrics. The third is described in Appendix \ref{app:pre_post_mobility}, concerning how each of the metrics behave when comparing pre- and post-disaster mobility.

We observe that all the mobility metrics we have studied produce resettlement decay curves that differ significantly for the control and IDP groups (Section \ref{sec:decays}, Figure \ref{fig:decay}), and that the results are significantly more different between the pre-disaster and post-disaster periods of the IDP group than the control group (Appendix \ref{app:pre_post_mobility}). This is consistent with what we would expect if our IDP detection method has correctly identified true IDPs. 

In terms of the potential advantages and disadvantages of each metric, which would benefit from further study, we make the following observations: the radius of gyration (RoG), as discussed in Section \ref{sec:metrics}, is problematic to use by itself because in many cases it is not sensitive to increases in short distance travel, and we have previously observed that many displacements occur over very short distances (\cite{li2019}). We saw in Section \ref{sec:decays} that the resettlement curves using RoG decayed much faster than the curves for the other metrics, for all three disasters, which may be because of the lack of sensitivity to short distance travel. It may also be because there are some behaviours that RoG is sensitive to, but not the other metrics. We have found also that the logarithmic radius of gyration (LRoG), temporal uncorrelated entropy (S\textsuperscript{uncorr}), and step entropy (S\textsuperscript{step}) all perform more strongly than the radius of gyration (RoG) with respect to the the two criteria described in the previous paragraph (different decay rates for the control and IDP groups, and a larger difference in behaviour between the pre- and post-disaster periods for the IDP group than control group). Of these three metrics, LRoG also has the advantage of being sensitive to distance. 

Further exploration of the similarities and differences between the metrics would be valuable in order to better understand the types of behaviours that each metric detects, and then to ascertain which metric gives the best approximation of the true resettlement time. 

\subsection{Comparison with previous studies}
In this section, we briefly compare our results with those obtained by previous studies of mobility in the aftermath of a disaster.

\subsubsection{Haiti earthquake, 2010 (\cite{lu2012})}
A study of the 2010 Haiti earthquake was performed by the authors of \cite{lu2012}, based on the same dataset as used in the current work, but using very different methods. The authors of \cite{lu2012} used the radius of gyration and entropy as mobility metrics, using the original definitions from \cite{song2010}. Several conclusions about post-disaster mobility were drawn, including: (i) after the disaster there was a large deviation in the distribution of travel distances, (ii) movement patterns for the affected people returned back to normal after four to five months, (iii) people moved farther, but not less regularly, in the aftermath of the disaster, and (iv) the movements of the population remained highly regular and predictable. 

The first conclusion (i) is consistent with what we have found and described in Section \ref{sec:decays} - that the values of all mobility metrics increased after the disaster, which indicates a deviation in travel distances (as well as travel frequency). The second conclusion (ii) is also consistent with our findings - the decay curves in Section \ref{sec:decays} indicate that most people had returned to normal after four to five months. However, for the third and fourth points, we have found that travel frequency increased in the aftermath of the disaster, which can be interpreted as a decrease in regularity and predictability. This highlights how the use of different metrics, and the consideration of different aspects of mobility, can lead to different conclusions about general mobility behaviour. 

\subsubsection{Nepal earthquake, 2015 (\cite{wilson2016})}
We compare the results that we obtained for the Nepal earthquake with those reported in \cite{wilson2016}. We will highlight that although the data used for the two studies is the same, the methodologies used are very different. The criteria used to identify someone as an IDP in \cite{wilson2016} is if they had spent at least seven consecutive days away from their pre-earthquake home location in a two week period after the earthquake, where home locations are determined as an administrative level 3 region. Return rates were determined by calculating the percentage of displaced users who remained away (at a different location to their pre-earthquake home location) on each date. In contrast, our method described in \cite{li2019} classified someone as an IDP if they had spent at least three consecutive days away from their home location, starting in the one week period after the disaster. Home locations were determined at the level of cell tower locations, or the centroid of a cluster (described in Section 3 of \cite{li2019}), so that people can be classified as displaced even if they remain within the same administrative level 3 region. Return rates were calculated as described in the Section \ref{sec:method}. In view of these differences, it can therefore be informative to make a qualitative comparison between the two studies, but not a quantitative one.

The authors of \cite{wilson2016} studied several administrative level 3 regions surrounding the earthquake epicentre in the Kathmandu Valley. It was found that $> 35\%$ of IDPs from each region were still displaced two weeks after the earthquake, and between 13-33\% from each region were still displaced five to six weeks after the earthquake. After 13 weeks there were 7-15\% remaining.

\begin{centering}
\begin{figure}
  \begin{subfigure}{1.0\linewidth}
    \centering\includegraphics[width=0.65\linewidth]{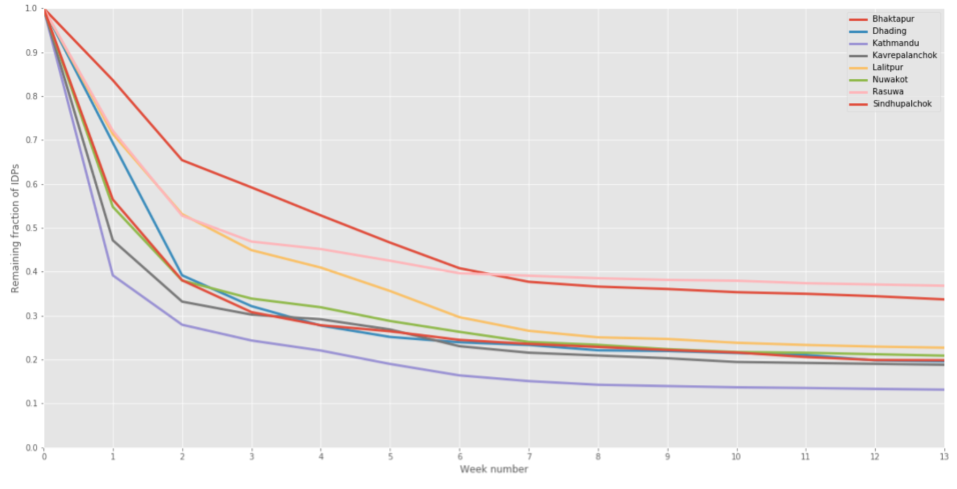}
    \caption{Resettlement decay curves when using the time of first return to the administrative level 3 region as the resettlement time.}
  \label{fig:bagmati_1}
  \end{subfigure}
  \begin{subfigure}{1.0\linewidth}
    \centering\includegraphics[width=0.65\linewidth]{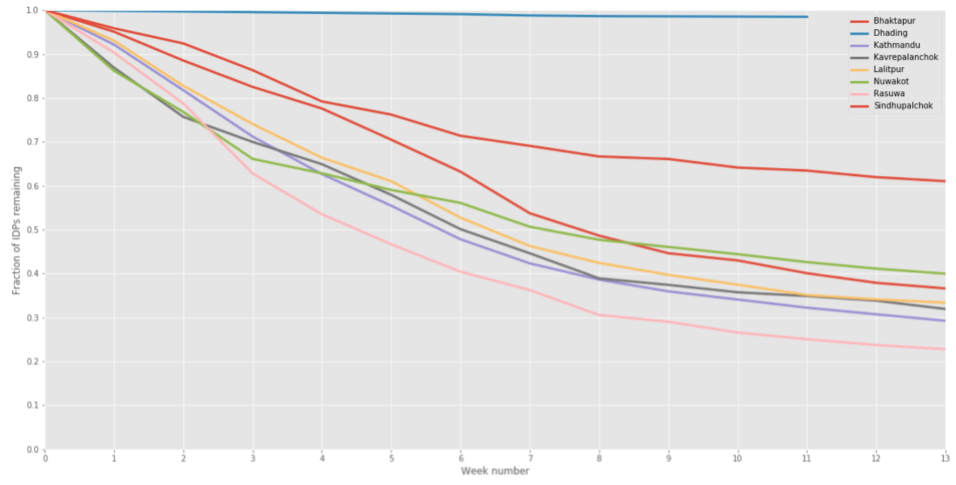}
    \caption{Resettlement decay curves when using the resettlement times computed in this work.}
  \label{fig:bagmati_2}
  \end{subfigure}
\caption{Resettlement decay curves for regions in Bagmati when using different methods to determine return time.}
\label{fig:bagmati}
\end{figure}
\end{centering}

In Figure \ref{fig:bagmati} we show the results obtained from our methods for the administrative level 3 regions in Bagmati which are included in \cite{wilson2016}. Figure \ref{fig:bagmati_1} shows the decay curves if someone is considered to have resettled on the first date that they returned to their administrative level 3 home region. This shows that after two weeks, there were $> 27\%$ of IDPs remaining, after six weeks there were 15-40\% remaining, and after 13 weeks there were 13-37\% remaining. Broadly speaking, these numbers are reasonably consistent with those in \cite{wilson2016} although the specific results for each region are quite different. In contrast, in Figure \ref{fig:bagmati_2} we show the decay curves corresponding to the definition of resettlement that has been used throughout this work and that is described in Section \ref{sec:method}. These recovery rates are much slower than those shown in Figure \ref{fig:bagmati_1} indicating that the time it takes people to truly `recover’ and return to their normal pre-disaster behaviour, as computed in the present work, is much longer than the time it may take them to physically return to their home location. This may be because the initial return to home may only be temporary and be for a number of purposes such as collecting belongings, assessing damage, or helping friends and family members.

\subsubsection{Hurricane Katrina, New Orleans, 2005 (\cite{fussell2010})}
In order to better understand the drivers of resettlement rates, it would be valuable to study a large number of disasters and their associated resettlement rates. Here, we briefly examine one work that studied a disaster that is different to the ones that we have studied in the current work.

The authors of \cite{fussell2010} studied the return of New Orleans residents who had been displaced by Hurricane Katrina, over a 14-month period, using data from a representative sample of residents collected in the Displaced New Orleans Residents Pilot Survey. The authors report that a quarter of displaced residents had returned to live in the city within two months, half within seven months, and slightly more than half after 14 months. This is a much slower recovery rate than what we have observed for all three of the disasters in our own work, where we find that at least half of the displaced residents have resettled back at their home after four months (see Table \ref{tab:halflife}). There are a large number of reasons that could explain the differences (aside from the differences in the data and methods used) such as the differences in the social, economic, political, and cultural contexts. Exploring these differences, and others, can potentially lead to insights regarding the ways in which resettlement times can be reduced through improved disaster preparedness or disaster response.

\section{Summary}

We have continued our work in \cite{li2019} and developed a method to analyse mobile phone call details records (CDRs) to estimate the time at which a person who has been displaced after a disaster returns to their normal pre-disaster behaviour. This is done by identifying the date at which each individual's mobility, after the disaster, returns to the same level as their `normal' mobility before the disaster. We assume that this corresponds to the end of their `displaced' situation and therefore interpret it as `resettlement', either at their original home location or at a new location. However, individuals who appear to have recovered normal activity may actually still be displaced, e.g. living in displacement sites or temporary accommodation. Therefore, we cannot measure with certainty the duration of displacements. However, our method still allows for a measurement of the recovery of the system and a return to normal activity.

We have presented the recovery rates in the form of `decay curves’ which show the remaining fraction of people who are still displaced at each point in time after the disaster. We find that these decay curves can be modelled extremely well as the sum of two exponential functions, which we interpret as corresponding to a fast decay (fast recovery) group of internally displaced persons, and a group that recovers more slowly. In further work we would like to explore what determines the values of these parameters, and therefore the extent to which these parameters can be predicted for future disaster scenarios. Related to this, we have also found that the resettlement rates for two of the studied disasters are extremely similar and that the rate for the third one is fairly similar. 

In the second half of this work, we compared the four mobility metrics that have been used in this study and identified the importance of ensuring that travel frequency, as well as travel distance, is taken into account when studying post-disaster mobility. This is because many people’s disrupted behaviour manifests as a decrease in long-distance travel but an increase in short-distance travel. This behaviour needs to be taken into account in order to accurately understand the true extent of disruption on a population and the support that is required.

We recognise that the methods we have presented in this work have limitations and that the results we have presented should be interpreted with a number of caveats in mind. Firstly, the original dataset of call detail records described in \cite{li2019} only contains records for people that use a mobile phone and the data are therefore unlikely to be representative of the entire population. In particular, the youngest and oldest members of the population, and individuals in the lowest socio-economic strata will likely not be represented. Furthermore, we then apply numerous filters in \cite{li2019} to select out only a small fraction of the original dataset to study. The anonymous individuals that are included in this study therefore constitute a very small and possibly biased sample of the entire population and it should not automatically be assumed that the results presented here are representative of the entire population - further work is needed to ascertain the extent to which this may or may not be true. Secondly, it is difficult to obtain the relevant ground-truth data against which our results can be validated, and so this has not yet been done. In spite of these limitations, we believe that this work can contribute towards a better understanding of post-disaster resettlement behaviour. 

\section{Acknowledgements}
This work was performed as part of the project `Contributing to a better understanding of human mobility in crisis and enhancing linkages with citizen-driven assistance’, a collaborative project between the International Organisation for Migration (IOM), the Internal Displacement Monitoring Centre (IDMC), the Humanitarian Data Exchange (HDX, part of the United Nations Office for the Coordination of Humanitarian Affairs (OCHA)), and Flowminder, funded by the European Civil Protection and Humanitarian Aid Operations (ECHO). Access to the de-identified mobile operator data used in this work was generously provided by Digicel (Haiti) and NCell (Nepal). Processing of that data was performed whilst adhering to high levels of data governance and information security best practices.

\bibliographystyle{unsrt}  
\bibliography{references}  

\appendix
\section{Pre-disaster decay curves}
\label{app:decay}
\begin{centering}
\begin{figure}[h!]
  \begin{subfigure}{1.0\linewidth}
    \centering\includegraphics[width=0.65\linewidth]{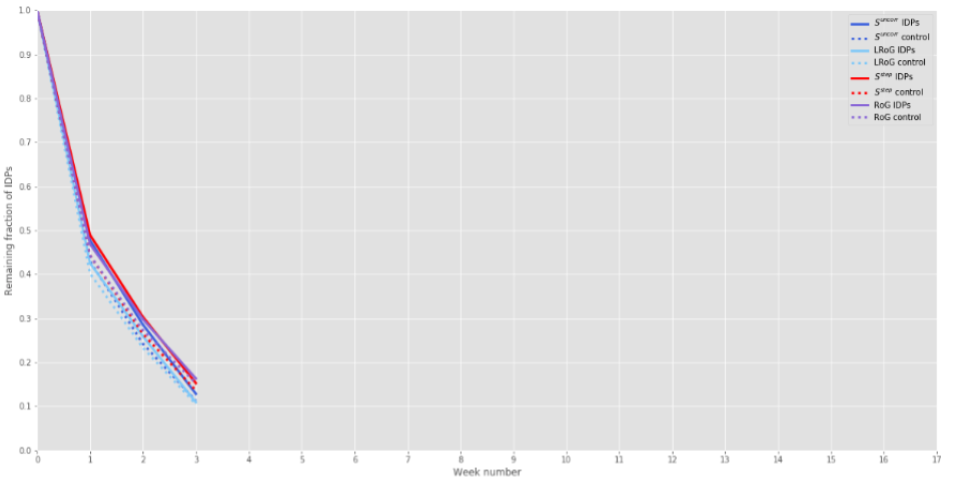}
    \caption{Haiti earthquake.}
  \end{subfigure}
  \begin{subfigure}{1.0\linewidth}
    \centering\includegraphics[width=0.65\linewidth]{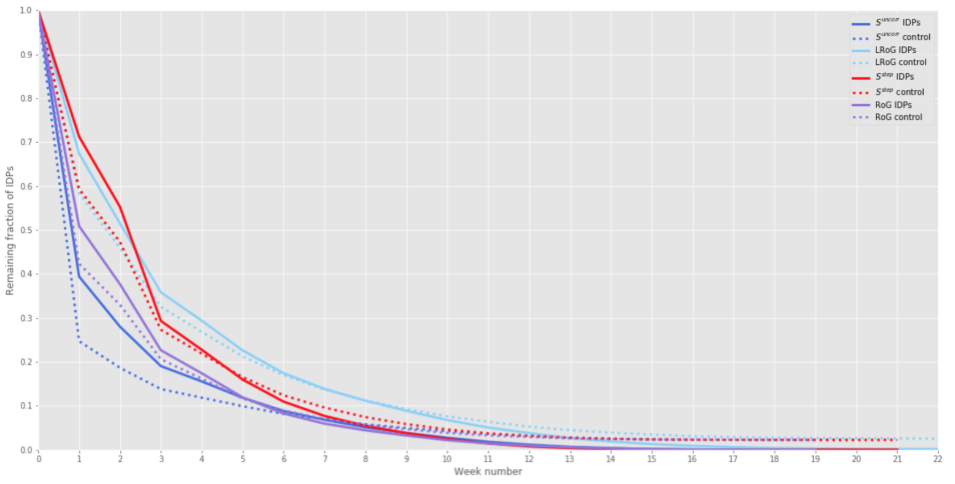}
    \caption{Hurricane Matthew.}
  \end{subfigure}
  \begin{subfigure}{1.0\linewidth}
    \centering\includegraphics[width=0.65\linewidth]{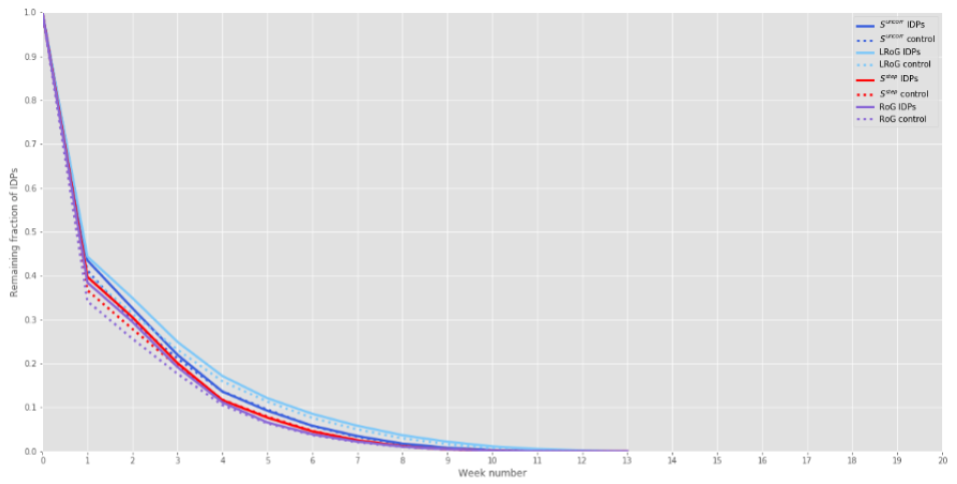}
    \caption{Nepal earthquake.}
  \end{subfigure}
\caption{Validation decay curves for the pre-disaster period. Results for the IDP group are shown by solid lines, and results for the control group are shown by dotted lines.}
\label{fig:predisaster}
\end{figure}
\end{centering}

As part of our verification that the method described in Section \ref{sec:method} to calculate recovery rates performs as expected, we calculated the `recovery rates’ for the pre-disaster period. These are shown in Figure \ref{fig:predisaster}. It should be expected that the control and IDP groups behave similarly during this period. We find that this is generally true, with the exception of weeks 1 to 4 of the Hurricane Matthew dataset, but there is still more similarity here than for the post-disaster period (Figure \ref{fig:decay}). The pre-disaster control groups do `recover’ slightly faster than the IDP groups though, for all disaster datasets and all metrics, indicating that the mobility behaviour of the IDPs is inherently more volatile than that of the control group.

\section{Comparing pre-disaster behaviour of control and IDP groups}
The z scores comparing the behaviours of the control and IDP groups in the pre-disaster period are shown in Table \ref{tab:predisaster}. These scores indicate that the two groups are significantly different even during the stable pre-disaster period. It is likely that this is because a certain behavioural subset has been selected by the filters included in the IDP detection method \cite{li2019}; for example, that an individual needed to have been seen at their home location during at least 90 percent of weeks in the pre-disaster period. RoG is consistently the least sensitive metric to these differences, for all the disasters. The smallest differences are observed in the Haiti earthquake dataset which may be a result of the sparseness of that dataset (only the location of the first call of the day is available - see \cite{li2019} for more details).

\begin{table}[h!]
\centering
\begin{tabular}{| l | l | l | l |}
  	\hline
	 	 & Haiti Earthquake & Hurricane Matthew & Nepal Earthquake \\
	 \hline
	 RoG & 8.23 & 34.7 & 37.1 \\
	 LRoG & 9.05 & 60.5 & 91.2 \\
	 S\textsuperscript{uncorr} & 29.1 & 54.1 & 86.9 \\
	 S\textsuperscript{step} & 35.2 & 60.1 & 77.6 \\
  	 \hline
\end{tabular}
\caption{z scores comparing the behaviours of the control and IDP groups during the pre-disaster period, for all four mobility metrics.}
\label{tab:predisaster}
\end{table}

\section{Distributions of mobility metrics in pre-disaster and post-disaster periods}
\label{app:pre_post_mobility}
\subsection{Comparing pre-disaster and post-disaster mobility}
Our hypothesis is that the mobility of people who are displaced by a disaster increases in the immediate aftermath of the disaster, where an increase in mobility means longer-distance travel and/or more frequent travel. To test this hypothesis, we calculate mobility metrics for each individual for the pre-disaster and post-disaster periods and compare the distributions obtained for these periods. We do this for both the control and IDP groups, for all four mobility metrics.

\begin{centering}
\begin{figure}[h!]
  \begin{subfigure}{1.0\linewidth}
    \includegraphics[width=0.5\linewidth]{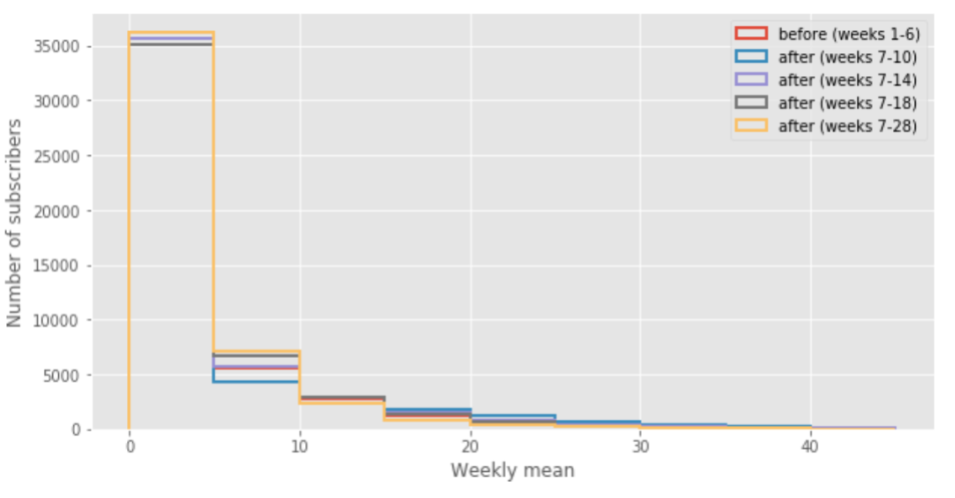}
    \includegraphics[width=0.5\linewidth]{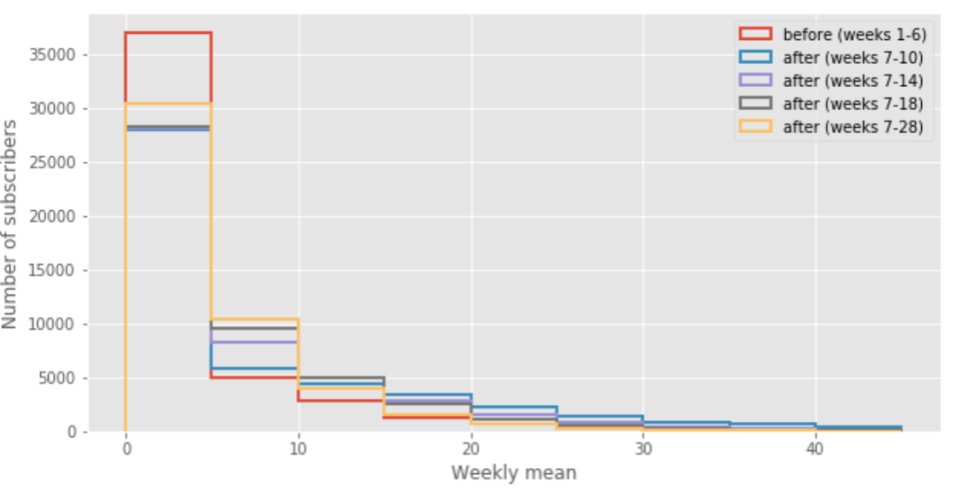}
    \caption{Haiti earthquake.}
  \end{subfigure} \\
  \begin{subfigure}{1.0\linewidth}
    \includegraphics[width=0.5\linewidth]{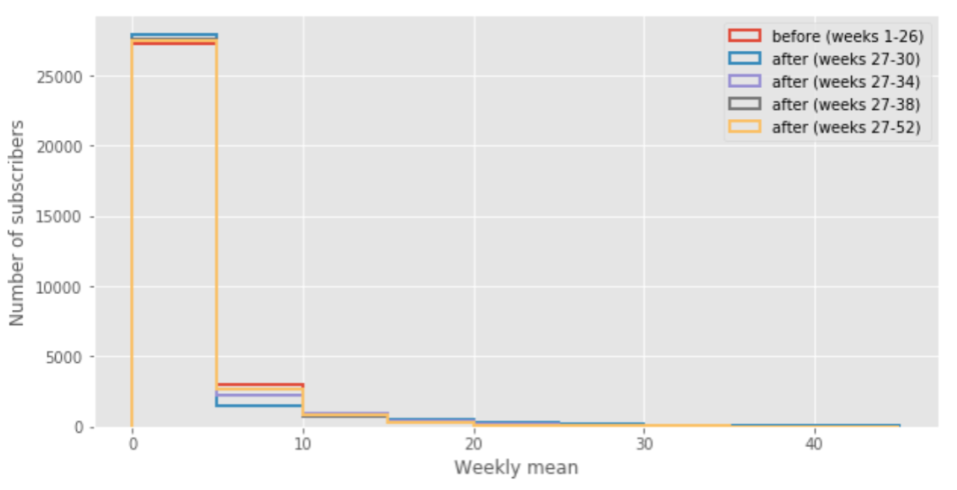}
    \includegraphics[width=0.5\linewidth]{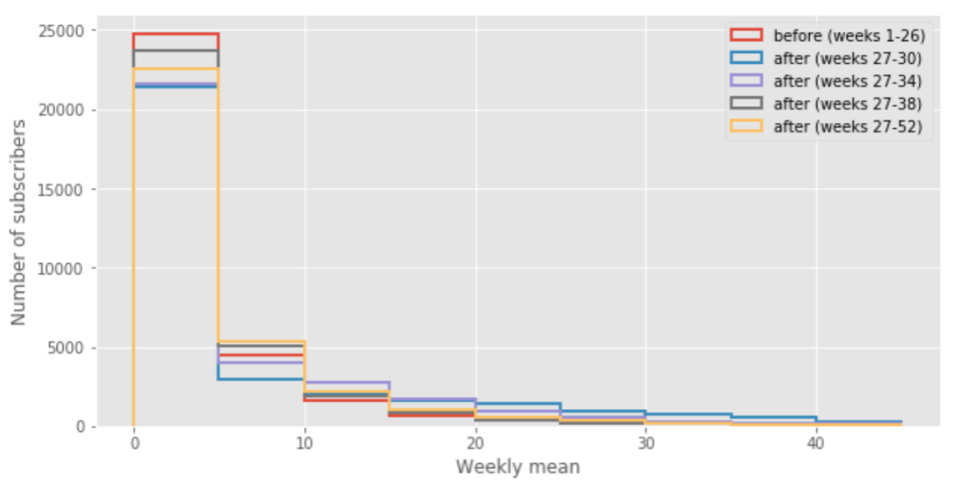}
    \caption{Hurricane Matthew.}
  \end{subfigure} \\
  \begin{subfigure}{1.0\linewidth}
    \includegraphics[width=0.5\linewidth]{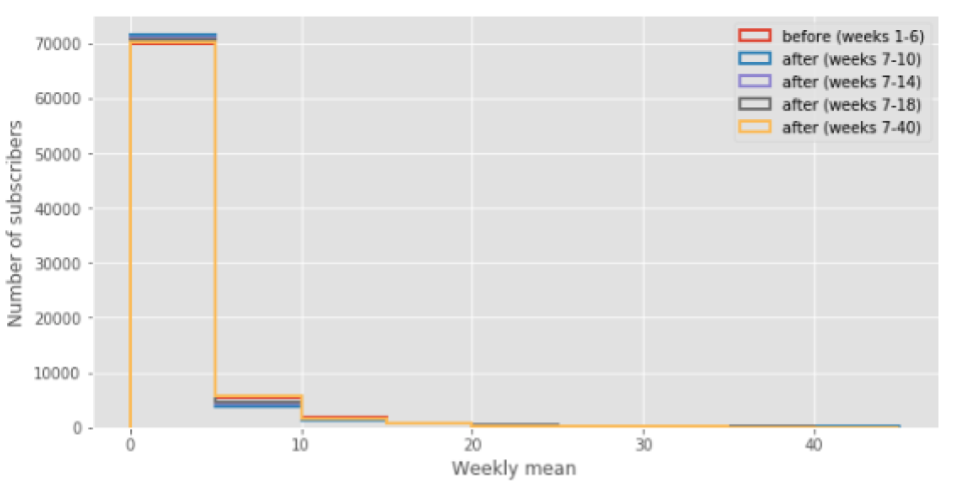}
    \includegraphics[width=0.5\linewidth]{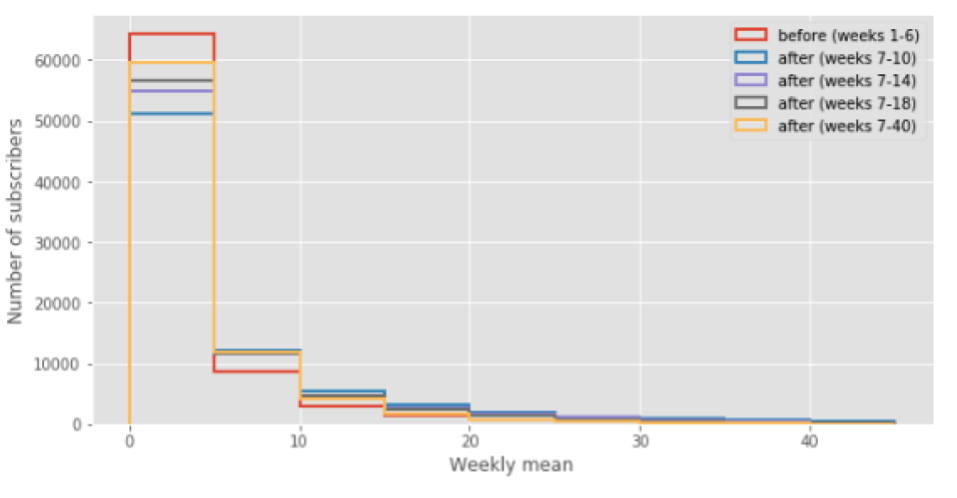}
    \caption{Nepal earthquake.}
  \end{subfigure}
\caption{Distributions of mean weekly RoG for different time periods, for control group (left column) and IDP group (right column).}
\label{fig:rog}
\end{figure}
\end{centering}

\begin{centering}
\begin{figure}[h!]
  \begin{subfigure}{1.0\linewidth}
    \includegraphics[width=0.5\linewidth]{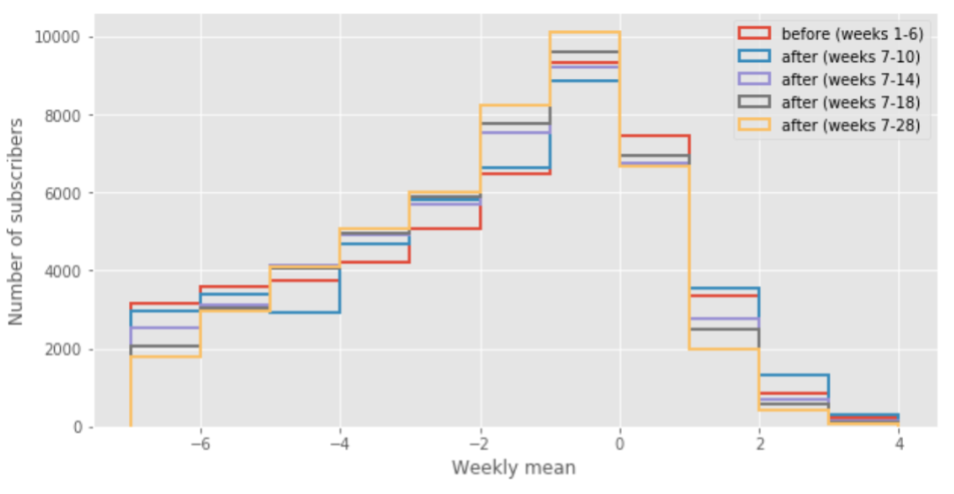}
    \includegraphics[width=0.5\linewidth]{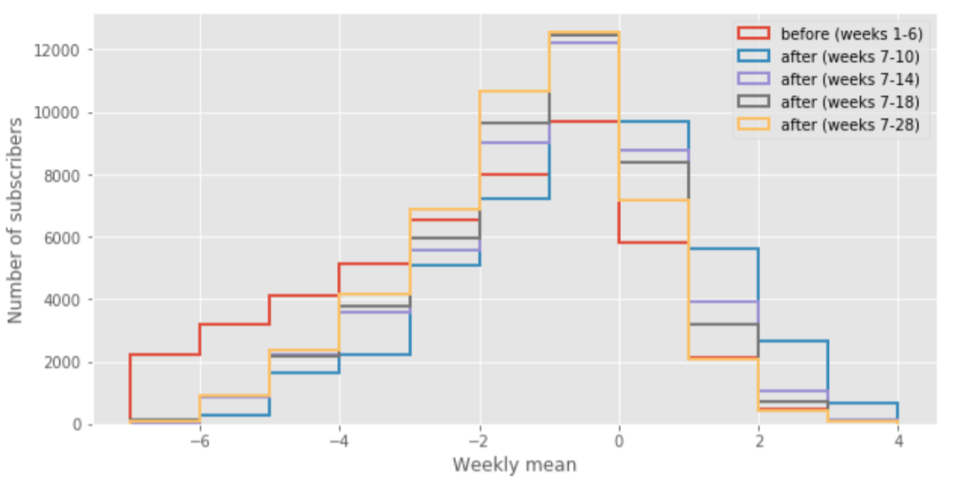}
    \caption{Haiti earthquake.}
  \end{subfigure} \\
  \begin{subfigure}{1.0\linewidth}
    \includegraphics[width=0.5\linewidth]{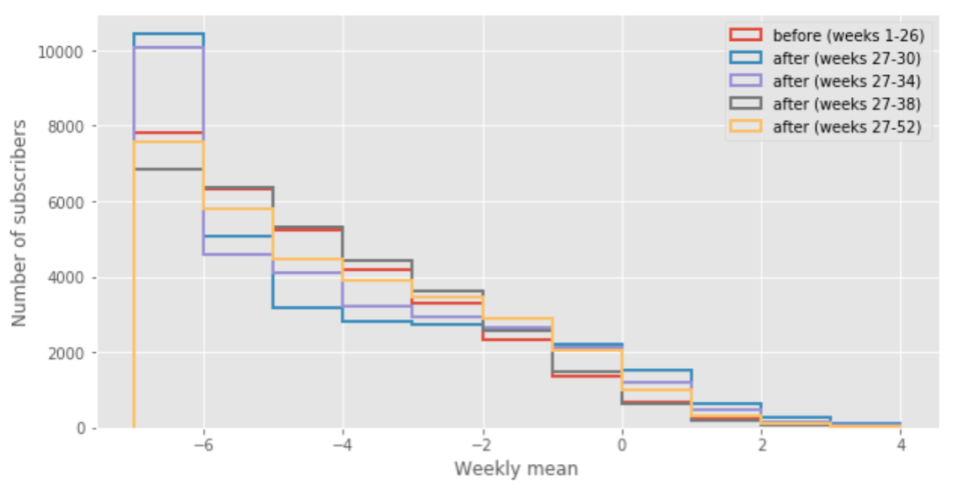}
    \includegraphics[width=0.5\linewidth]{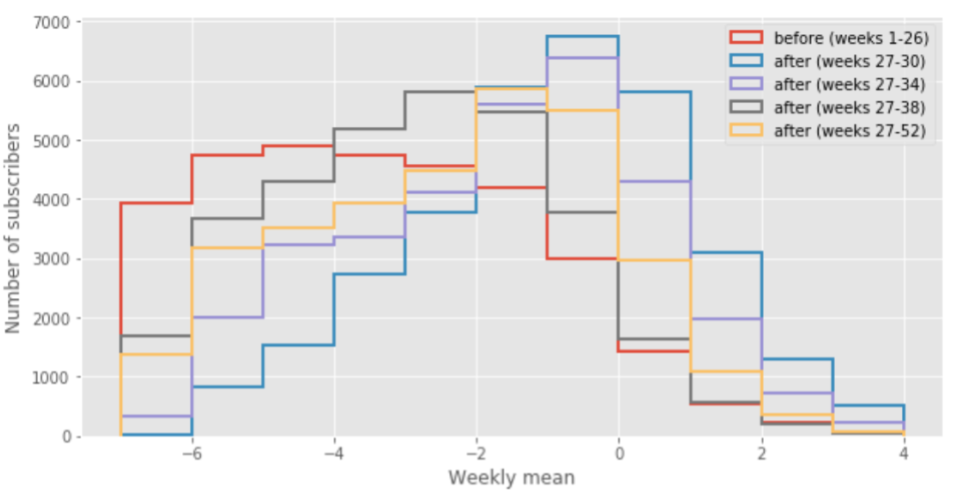}
    \caption{Hurricane Matthew.}
  \end{subfigure} \\
  \begin{subfigure}{1.0\linewidth}
    \includegraphics[width=0.5\linewidth]{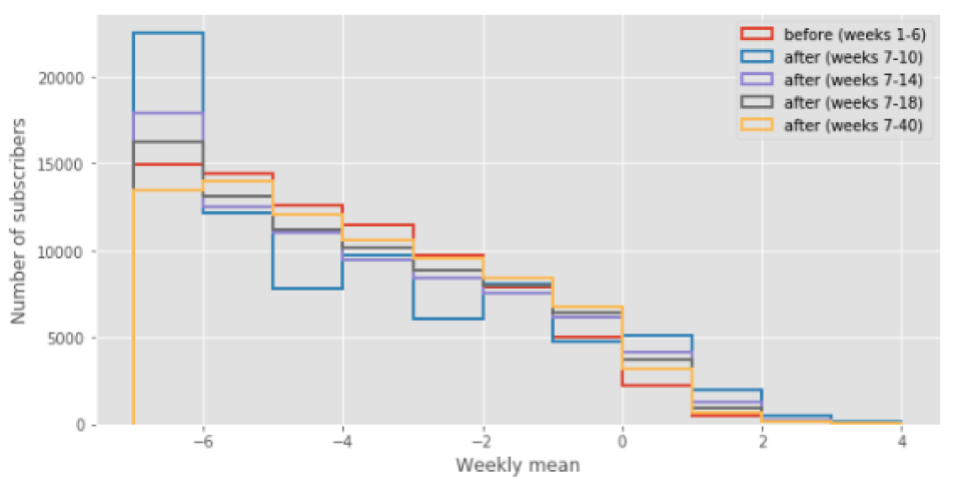}
    \includegraphics[width=0.5\linewidth]{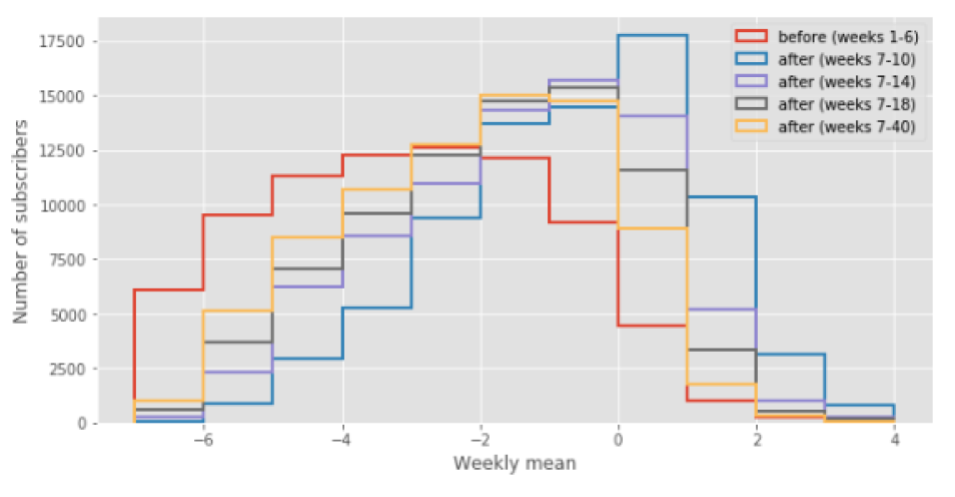}
    \caption{Nepal earthquake.}
  \end{subfigure}
\caption{Distributions of mean weekly LRoG for different time periods, for control group (left column) and IDP group (right column).}
\label{fig:lrog}
\end{figure}
\end{centering}

\begin{centering}
\begin{figure}[h!]
  \begin{subfigure}{1.0\linewidth}
    \includegraphics[width=0.5\linewidth]{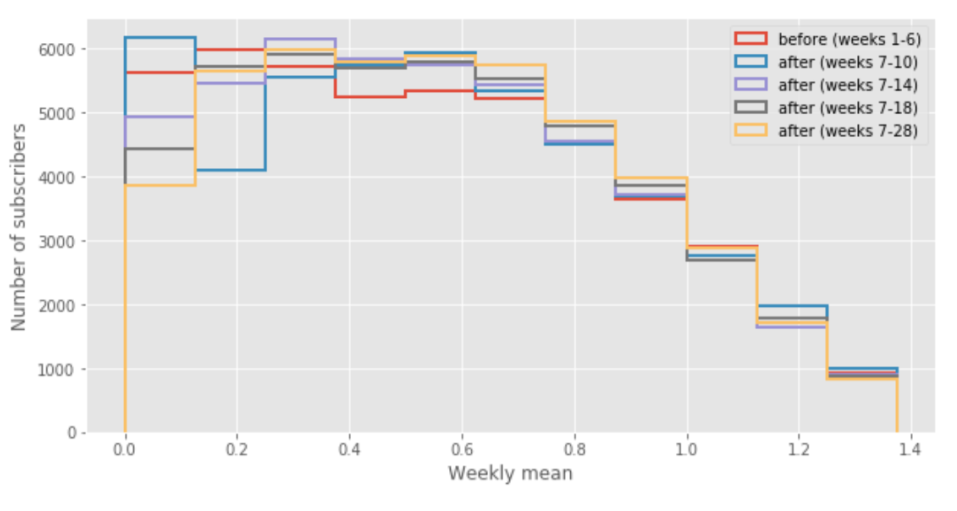}
    \includegraphics[width=0.5\linewidth]{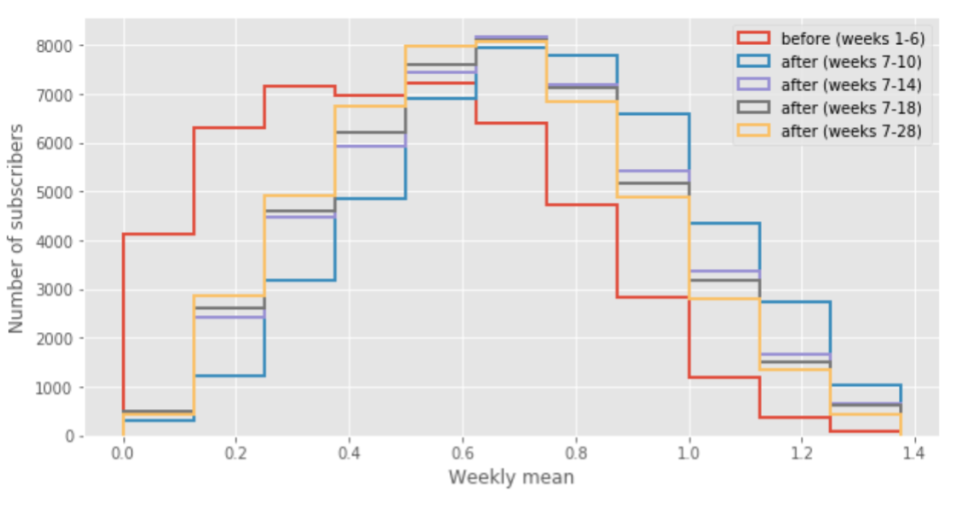}
    \caption{Haiti earthquake.}
  \end{subfigure} \\
  \begin{subfigure}{1.0\linewidth}
    \includegraphics[width=0.5\linewidth]{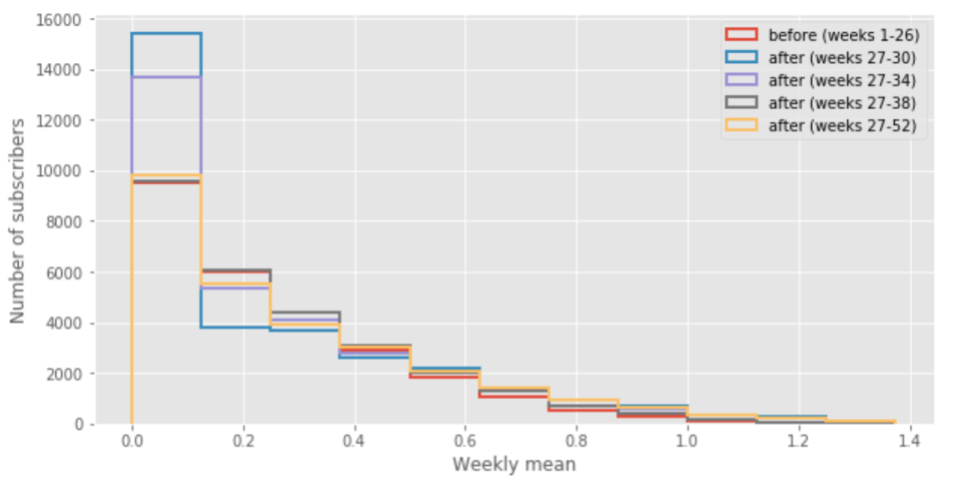}
    \includegraphics[width=0.5\linewidth]{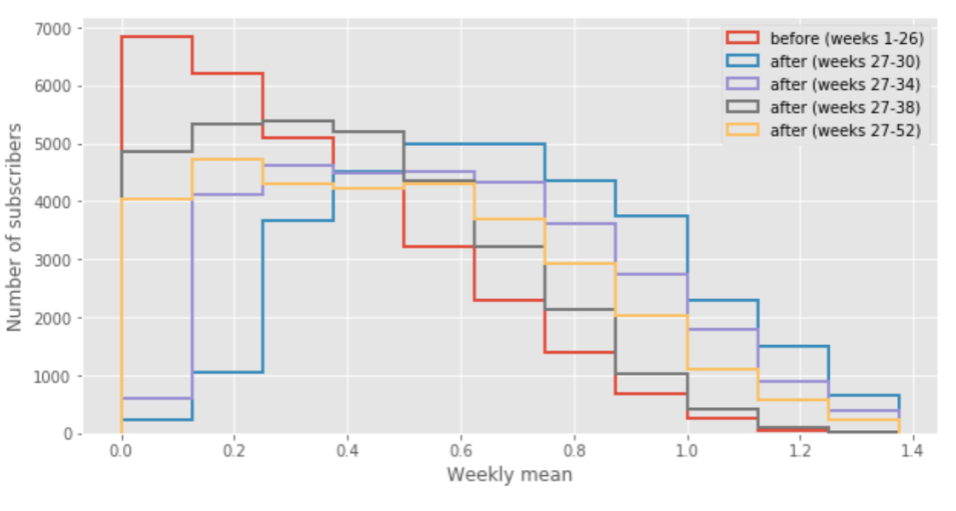}
    \caption{Hurricane Matthew.}
  \end{subfigure} \\
  \begin{subfigure}{1.0\linewidth}
    \includegraphics[width=0.5\linewidth]{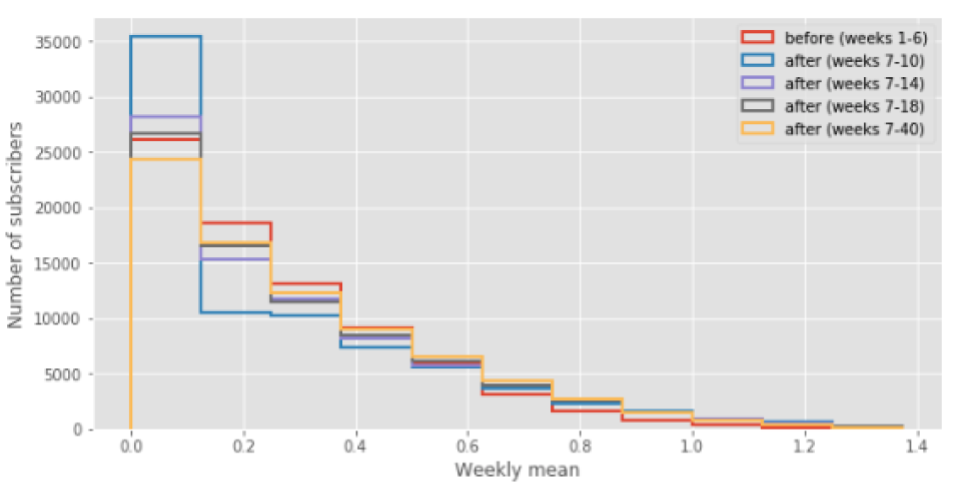}
    \includegraphics[width=0.5\linewidth]{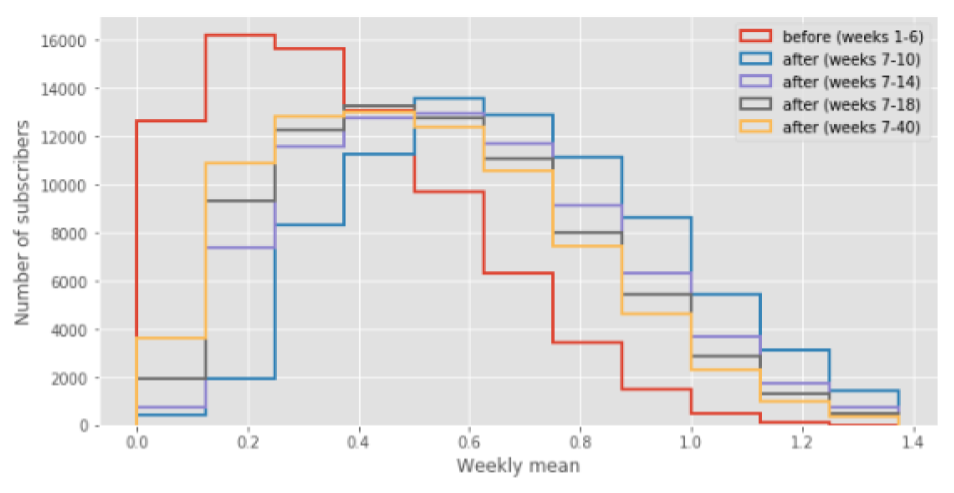}
    \caption{Nepal earthquake.}
  \end{subfigure}
\caption{Distributions of mean weekly S\textsuperscript{uncorr} for different time periods, for control group (left column) and IDP group (right column).}
\label{fig:suncorr}
\end{figure}
\end{centering}

\begin{centering}
\begin{figure}[h!]
  \begin{subfigure}{1.0\linewidth}
    \includegraphics[width=0.5\linewidth]{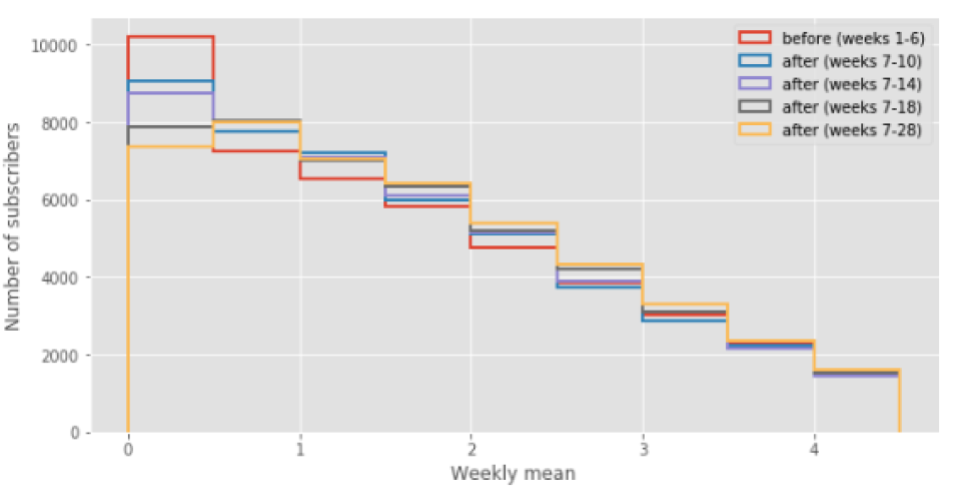}
    \includegraphics[width=0.5\linewidth]{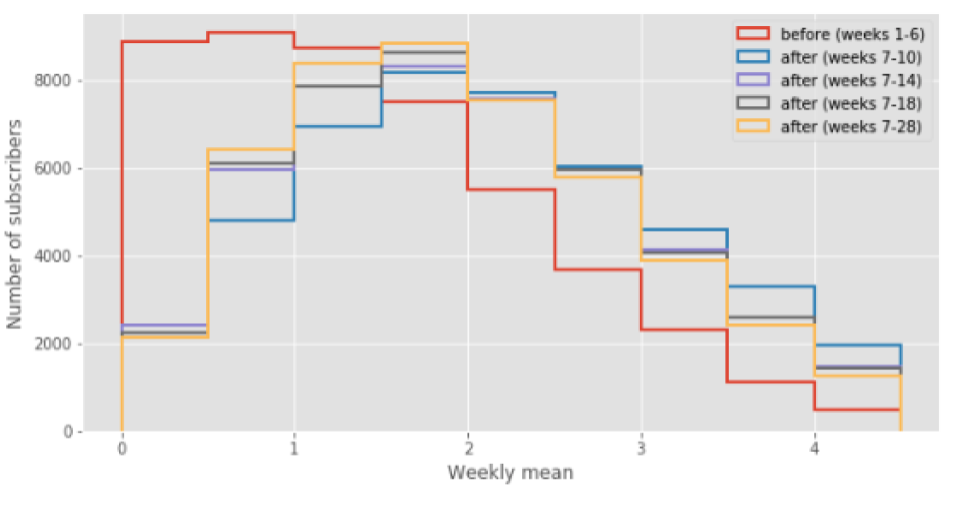}
    \caption{Haiti earthquake.}
  \end{subfigure} \\
  \begin{subfigure}{1.0\linewidth}
    \includegraphics[width=0.5\linewidth]{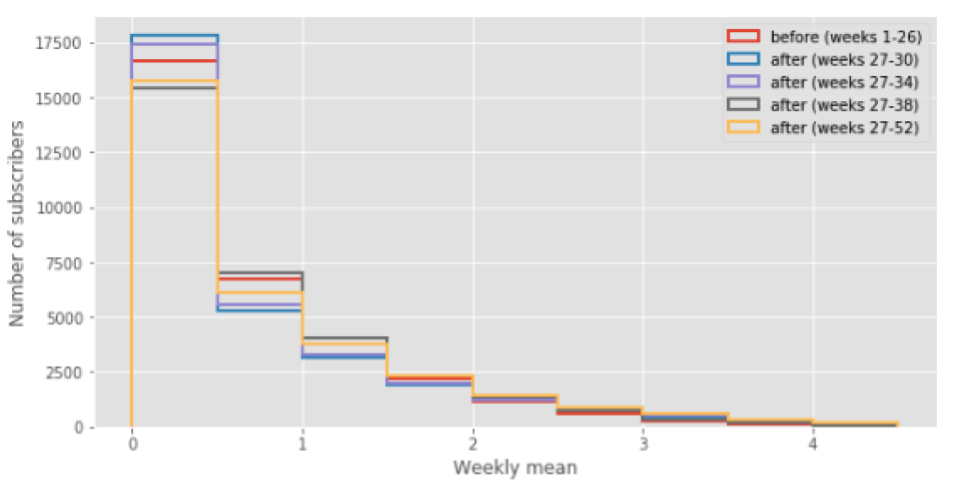}
    \includegraphics[width=0.5\linewidth]{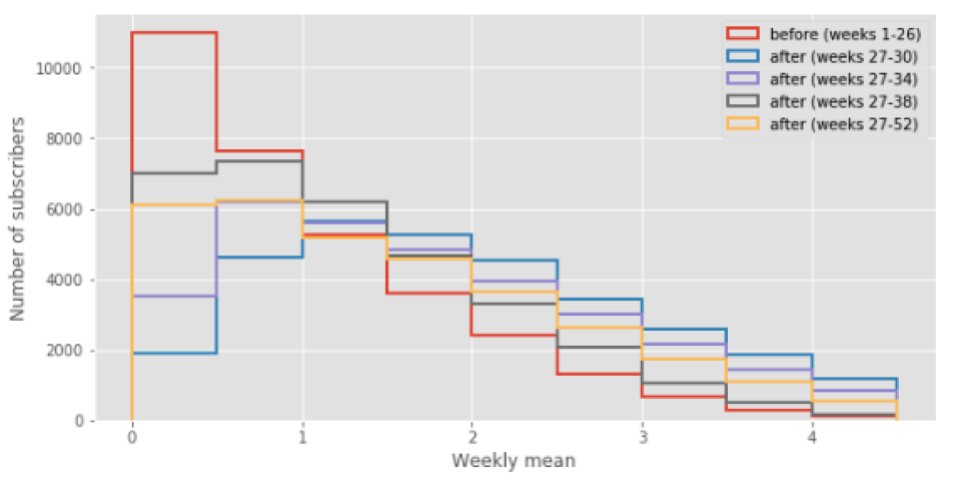}
    \caption{Hurricane Matthew.}
  \end{subfigure} \\
  \begin{subfigure}{1.0\linewidth}
    \includegraphics[width=0.5\linewidth]{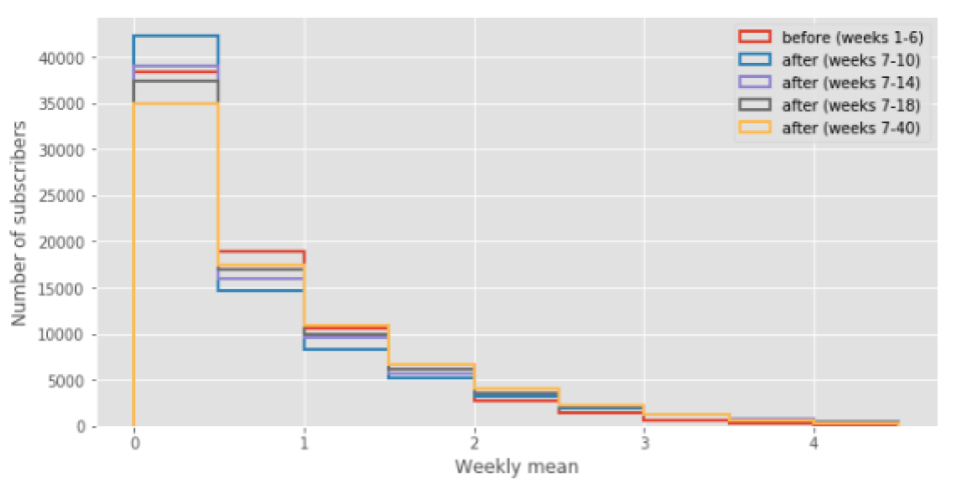}
    \includegraphics[width=0.5\linewidth]{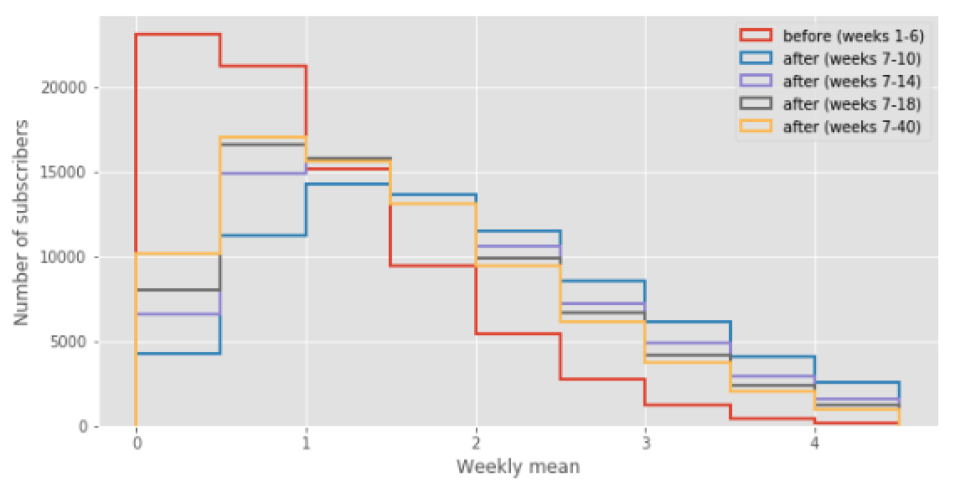}
    \caption{Nepal earthquake.}
  \end{subfigure}
\caption{Distributions of mean weekly S\textsuperscript{step} for different time periods, for control group (left column) and IDP group (right column).}
\label{fig:sstep}
\end{figure}
\end{centering}

We have compared how the distributions of each of the four mobility metrics differs between the pre-disaster period and several post-disaster time periods, for both the control and IDP groups. The mean weekly values of each of the metrics was calculated for the following periods: the pre-disaster period, the four-week period immediately following the disaster, the eight-week period immediately following the period, the 12-week period immediately following the disaster, and the complete post-disaster period which is of different durations for each disaster (these durations are shown in Table \ref{tab:disasters}). The distributions are shown in Figures \ref{fig:rog}, \ref{fig:lrog}, \ref{fig:suncorr}, and \ref{fig:sstep}. Comparing the different metrics, a visual inspection shows that when using RoG, there is no significant difference between the pre- and post-disaster distributions, and between the control and IDP groups. However, when using LRoG or either of the entropy metrics, there are very obvious visual differences between time periods, and between the control and IDP groups. 

\clearpage

Paired 2-sample z scores were calculated to test if there is a significant difference in means between the pre-disaster period and each of the post-disaster periods. These are shown in Tables \ref{tab:weekly_rog}, \ref{tab:weekly_lrog}, \ref{tab:weekly_suncorr}, and \ref{tab:weekly_sstep}. The z scores for the control group are, in most cases and for all the mobility metrics, high enough to indicate that the behaviour of this group is significantly different in the post-disaster period than in the pre-disaster period and that mobility increased after the disaster, even though we assume that the majority of individuals in this group were not affected by the disaster. This is likely due to one or more of the following reasons: a high proportion of individuals in this group are actually IDPs and thus were affected by the disaster, there are not many IDPs in the group but people’s behaviour is highly variable anyway, or the methods we have used are such that this variability has been introduced artificially. However, the z scores for the IDP group are much higher, relative to the scores for the control group, so that we can have confidence that the behaviour of the IDP group was affected more by the disaster, in terms of an increase in mobility, than that of the control group. This is consistent with our hypothesis. 

\begin{table}[h!]
\centering
\begin{tabular}{| l | l | l | l | l | l | l |}
  	\hline
	 & \multicolumn{2}{| c |}{Haiti Earthquake} & \multicolumn{2}{| c |}{Hurricane Matthew} & \multicolumn{2}{| c |}{Nepal Earthquake} \\
	 \hline
	 & Control & IDP & Control & IDP & Control & IDP \\
	 \hline
         Post-disaster - 4 weeks & 13.4 & 77.9 & 3.70 & 53.5 & 8.59 & 79.7 \\
	 Post-disaster - 8 weeks & 4.35 & 61.7 & 5.79 & 37.2 & 5.01 & 60.5 \\
 	 Post-disaster - 12 weeks & 3.43 & 54.8 & 0.45 & 11.4 & 4.44 & 46.9 \\
	 Post-disaster - all weeks & 5.43 & 35.7 & 0.87 & 21.3 & 4.66 & 25.3 \\
  	\hline
\end{tabular}
\caption{z values comparing mean weekly RoG from pre- and post-disaster periods, for control and IDP groups.}
\label{tab:weekly_rog}
\end{table}

\begin{table}[h!]
\centering
\begin{tabular}{| l | l | l | l | l | l | l |}
  	\hline
	 & \multicolumn{2}{| c |}{Haiti Earthquake} & \multicolumn{2}{| c |}{Hurricane Matthew} & \multicolumn{2}{| c |}{Nepal Earthquake} \\
	 \hline
	 & Control & IDP & Control & IDP & Control & IDP \\
	 \hline
         Post-disaster - 4 weeks & 6.19 & 117 & 5.64 & 141 & 1.57 & 236 \\
	 Post-disaster - 8 weeks & 1.89 & 77.9 & 5.61 & 93.7 & 11.5 & 159 \\
 	 Post-disaster - 12 weeks & 0.99 & 70.0 & 7.53 & 29.8 & 14.6 &126 \\
	 Post-disaster - all weeks & 0.16 & 55.2 & 14.3 & 57.6 & 20.6 & 91.5 \\
  	\hline
\end{tabular}
\caption{z values comparing mean weekly LRoG from pre- and post-disaster periods, for control and IDP groups.}
\label{tab:weekly_lrog}
\end{table}

\begin{table}[h!]
\centering
\begin{tabular}{| l | l | l | l | l | l | l |}
  	\hline
	 & \multicolumn{2}{| c |}{Haiti Earthquake} & \multicolumn{2}{| c |}{Hurricane Matthew} & \multicolumn{2}{| c |}{Nepal Earthquake} \\
	 \hline
	 & Control & IDP & Control & IDP & Control & IDP \\
	 \hline
         Post-disaster - 4 weeks & 6.35 & 138 & 31.2 &165 & 2.80 & 248 \\
	 Post-disaster - 8 weeks & 1.89 & 77.9 & 5.61 & 93.7 & 17.2 & 176 \\
 	 Post-disaster - 12 weeks & 5.59 & 94.6 & 16.3 & 39.8 & 21.2 &146 \\
	 Post-disaster - all weeks & 8.76 & 85.4 & 30.4 & 73.6 & 31.2 &119 \\
  	 \hline
\end{tabular}
\caption{z values comparing mean weekly S\textsuperscript{uncorr} from pre- and post-disaster periods, for control and IDP groups.}
\label{tab:weekly_suncorr}
\end{table}

\begin{table}[h!]
\centering
\begin{tabular}{| l | l | l | l | l | l | l |}
  	\hline
	 & \multicolumn{2}{| c |}{Haiti Earthquake} & \multicolumn{2}{| c |}{Hurricane Matthew} & \multicolumn{2}{| c |}{Nepal Earthquake} \\
	 \hline
	 & Control & IDP & Control & IDP & Control & IDP \\
	 \hline
         Post-disaster - 4 weeks & 0.674 & 108 & 5.59 &121 & 6.43 & 200 \\
	 Post-disaster - 8 weeks & 0.672 & 87.7 & 8.06 & 89.8 & 20.7 &155 \\
 	 Post-disaster - 12 weeks & 5.27 & 86.7 & 11.2 & 36.8 & 25.0 &134 \\
	 Post-disaster - all weeks & 7.81 & 80.1 & 20.1 & 65.0 & 35.1 &115 \\
	 \hline
\end{tabular}
\caption{z values comparing mean weekly S\textsuperscript{step} from pre- and post-disaster periods, for control and IDP groups.}
\label{tab:weekly_sstep}
\end{table}

Interpreting the z values to indicate the magnitude of the effect size (the amount by which mobility increased after the disaster), in Table \ref{tab:trends} we have summarised the trends over time that are indicated by the z values, starting from the shortest time period after the disaster (four weeks) and finishing with the complete post-disaster period. The pattern that we would expect to observe from the control groups, who we expect to have ‘normal’ behaviour throughout the entire period, would be that behaviour becomes increasingly different as time progresses, such that someone’s behaviour is almost identical to their behaviour from the most recent time period, and different from their behaviour a long time ago. For the IDP groups, we expect the greatest effect to be immediately after the disaster, and for behaviour to return to normal - so the effect size to decrease - as time progresses. We find that the S\textsuperscript{step} metric best indicates these behaviours across all the disasters, and that LRoG is mostly consistent except for the control group from the Haiti earthquake. The other two metrics - RoG and S\textsuperscript{uncorr} - are not generally consistent with this pattern.

\begin{table}
  	\centering
  	\begin{tabular}{| l | l | l | l | l | l | l |}
  	\hline
	 & \multicolumn{2}{| c |}{Haiti Earthquake} & \multicolumn{2}{| c |}{Hurricane Matthew} & \multicolumn{2}{| c |}{Nepal Earthquake} \\
	 \hline
	 & Control & IDP & Control & IDP & Control & IDP \\
	 \hline
 	RoG & decreases & decreases & erratic & erratic & decreases & decreases \\
	LRoG & decreases & decreases & increases & decreases & increases & decreases \\
	S\textsuperscript{uncorr} & erratic & erratic & erratic & erratic & increases & decreases \\
	S\textsuperscript{step} & increases & decreases & increases & decreases & increases & decreases \\
 	\hline
  	\end{tabular}
	\caption{Trends over time in the effect size detected by different mobility metrics during  the post-disaster duration.}
  	\label{tab:trends}
\end{table}

\end{document}